\documentclass[aps,prl,twocolumn,amsmath,longbibliography,superscriptaddress]{revtex4-2}  % for review and submission
\usepackage{bbm}
\usepackage{mathrsfs}
\usepackage{amsmath}
\usepackage{amsfonts}
\usepackage[colorlinks=true,citecolor=blue,anchorcolor=blue]{hyperref}
\usepackage{graphicx,epstopdf}
\usepackage{subfigure}
\usepackage{epsfig}
\usepackage{dcolumn}
\usepackage{bm}
\usepackage{color}
\usepackage{natbib}
\usepackage{amssymb}
\usepackage{xcolor}
\usepackage{braket}

\begin{document}

\title{Floquet odd-parity collinear magnets}

% \author{Tongshuai Zhu$^{1,2}$, Di Zhou$^3$, Huaiqiang Wang$^{4,*}$, Jiawei Ruan$^{3,\dagger}$}

% \affiliation{
% 	$^1$ College of Science, China University of Petroleum (East China), Qingdao 266580, China\\
% $^2$ School of Materials Science and Engineering, China University of Petroleum (East China), Qingdao 266580, China\\
% $^3$ Eastern Institute of Technology, Ningbo 315200, China \\
% $^4$ Center for Quantum Transport and Thermal Energy Science, School of Physics and Technology, Nanjing Normal University, Nanjing 210023, China \\}

\author{Tongshuai Zhu}
\affiliation{College of Science, China University of Petroleum (East China), Qingdao 266580, China}
\author{Di Zhou}
\affiliation{Eastern Institute of Technology, Ningbo 315200, China}
\author{Huaiqiang Wang}
\email{hqwang@njnu.edu.cn}
\affiliation{Center for Quantum Transport and Thermal Energy Science, Institute of Physics Frontiers and Interdisciplinary Sciences, School of Physics and Technology, Nanjing Normal University, Nanjing 210023, China}
\affiliation{National Laboratory of Solid State Microstructures, Nanjing University, Nanjing 210093, China}
\affiliation{Jiangsu Physical Science Research Center, Nanjing 210093, China}

\author{Su-Huai Wei}
\affiliation{Eastern Institute of Technology, Ningbo 315200, China}

\author{Jiawei Ruan}
\email{jwruan@eitech.edu.cn}
\affiliation{Eastern Institute of Technology, Ningbo 315200, China}

\begin{abstract}
Altermagnets (AMs), recently discovered unconventional magnets distinct from both ferro- and antiferromagnets, have rapidly emerged as a prominent research topic in condensed matter physics. AMs are characterized by alternating collinear magnetic moments with zero net magnetization in real space, and spin splittings with even-parity symmetry in momentum space. However, their counterparts exhibiting odd-parity spin splittings are generally thought to be absent in collinear magnets. Here, we show that such unconventional odd-parity magnets can be induced from collinear antiferromagnets by symmetry engineering. Remarkably, using effective model analysis within Floquet-theory framework, we demonstrate that circularly polarized light irradiation of conventional antiferromagnetic lattices breaks a spin-preserving pseudo-time-reversal symmetry and induces both $p$- and $f$-wave magnets, realizing novel magnetic states dubbed \emph{Floquet odd-parity collinear magnets}. Moreover, we also uncover light-induced antiferromagnetic Chern insulating states in the $f$-wave magnets. The proposed Floquet odd-parity magnet is confirmed by first-principles calculations of MnPSe$_{3}$ under circularly polarized light. Our work not only proposes a new class of unconventional magnets, but also opens an avenue for light-induced magnetic phenomena in spintronic applications.

\end{abstract}

\maketitle

\emph{Introduction.} Unconventional magnets beyond ferromagnets (FMs) and antiferromagnets (AFMs), particularly the recently discovered altermagnet (AM), have attracted significant attention in condensed matter physics~\cite{vsmejkal2022emerging,bai2024altermagnetism,song2025altermagnets,vsmejkal2022beyond,wu2007fermi,ma2021multifunctional,mazin2023induced,zeng2024description,regmi2025altermagnetism,wang2024electric,Liu2024Twisted,duan2025antiferroelectric,Leeb2024Spontaneous,Zhu2025Design,Gu2025Ferroelectric,reimers2024direct,zhou2025manipulation,Karube2022Observation,bose2022tilted,fedchenko2024observation,Ding2024Large,han2024electrical,reichlova2024observation,lee2024broken,Osumi2024Observation,krempasky2024altermagnetic,Mazin2023altermagnetism,Hariki2024X-Ray,jiang2025metallic,zhang2025crystal,Banerjee2024Altermagnetic,vsmejkal2023chiral,xu2025altermagnetic,Che2025Engineering,gao2025ai,Pan2024General,Lin2025Coulomb,Chen2025Electrical,Ghorashi2025Altermagnetic,Yershov2025Curvature,Ouassou2023dc,Yuan2020Giant,Mondal2025Distinguishing}. AMs exhibit compensated collinear magnetic moments characteristic of AFMs in real space, while manifesting spin-splitting effects akin to FMs in reciprocal space. A diverse range of altermagnetic materials has been theoretically predicted~\cite{ma2021multifunctional,mazin2023induced,zeng2024description,wang2024electric,Liu2024Twisted,duan2025antiferroelectric,Leeb2024Spontaneous,Zhu2025Design,Gu2025Ferroelectric,xu2025altermagnetic,Che2025Engineering,gao2025ai,Pan2024General,Lin2025Coulomb,Chen2025Electrical,Ghorashi2025Altermagnetic,Yershov2025Curvature,Ouassou2023dc}, and experimental confirmation has already been achieved for some of their unconventional electronic and magnetic signatures~\cite{reimers2024direct,zhou2025manipulation,Karube2022Observation,bose2022tilted,fedchenko2024observation,Ding2024Large,han2024electrical,reichlova2024observation, lee2024broken,Osumi2024Observation,krempasky2024altermagnetic,Mazin2023altermagnetism,Hariki2024X-Ray,jiang2025metallic,zhang2025crystal,regmi2025altermagnetism}. Importantly, spin splittings in the intensively studied AMs typically exhibit even-parity symmetry in momentum space, manifested by Fermi surfaces featuring $d$-, $g$-, and $i$-wave symmetries~\cite{vsmejkal2022emerging, vsmejkal2022beyond}. In contrast, magnets exhibiting unconventional odd-parity spin splittings, such as those with $p$- and $f$-wave symmetries, remain largely unexplored.  Although odd-parity magnets have been proposed in noncollinear magnets~\cite{brekke2024minimal, yu2025odd, song2025electrical, hellenes2023p,Hayami2020Spontaneous,yamada2025metallic}, their complex magnetic structures are usually susceptible to external fields, thereby severely limiting prospects for spintronic applications. Consequently, realizing odd-parity magnets in collinear magnetic systems is essential, yet remains a critical challenge.

\begin{figure}[htbp]
	\centering
	\includegraphics[width=3.5in]{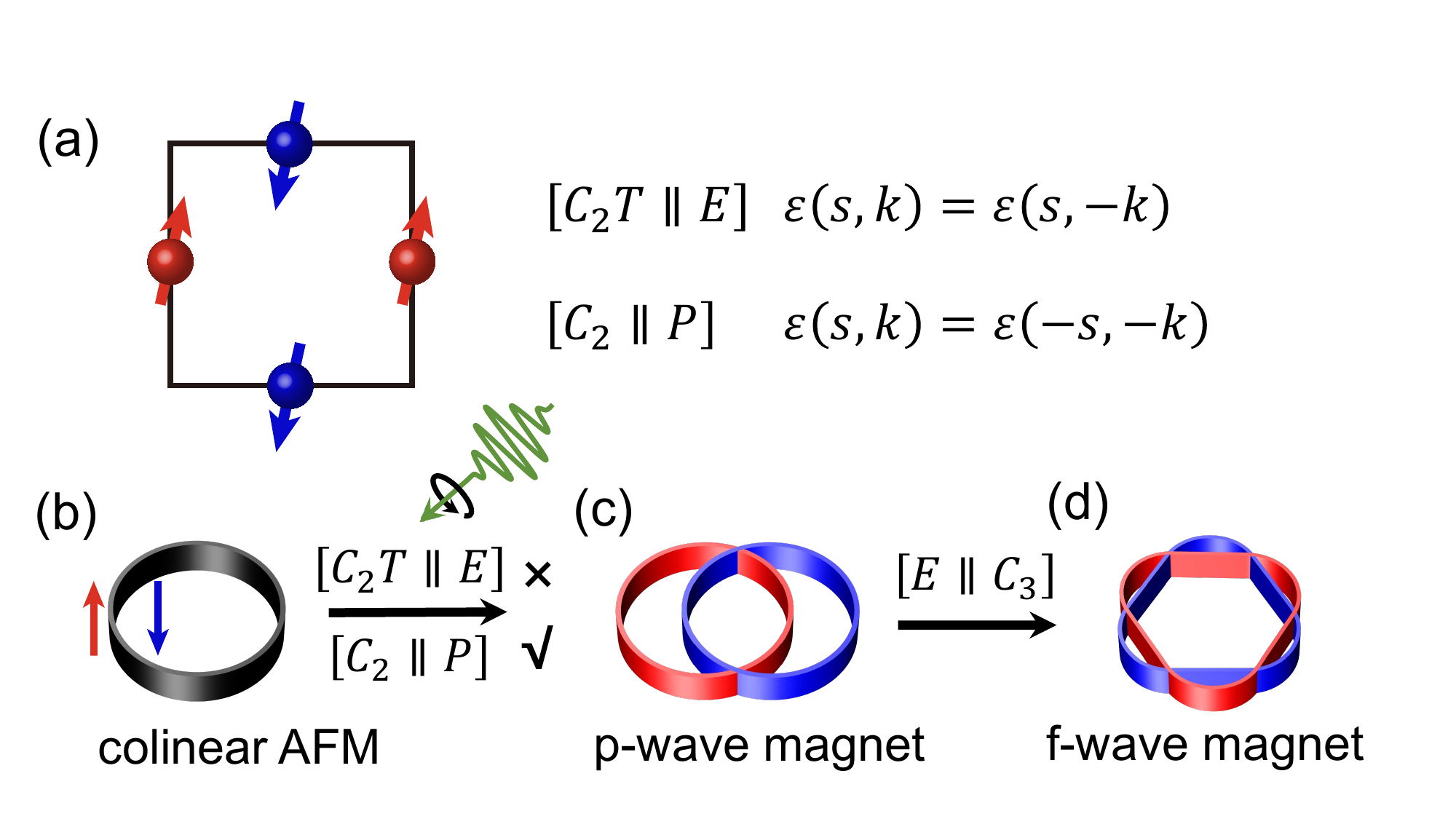}
	\caption{Schematic of odd-parity magnetism induced from collinear antiferromagnets (AFMs). (a) In conventional collinear $PT$-symmetric AFMs with negligible spin-orbit coupling, the two spin-group symmetries  $[C_{2}T||E]$ and $[C_{2}||P]$ connecting opposite momenta enforce spin degenerate bands. (b) Spin degenerate Fermi surface of the parent AFM. (c-d) Typical spin-polarized Fermi surfaces for the odd-parity magnets with (c) the p-wave magnet induced from the parent AFM by breaking the $[C_{2}T||E]$ symmetry while preserving the $[C_{2}||P]$ symmetry (e.g., by CPL) and (d) the f-wave magnet when the $p$-wave magnet possesses an additional three-fold rotation symmetry.}
	\label{fig1}
\end{figure}

 To achieve unconventional odd-parity spin splittings, a rational design principle involves manipulating electronic structures of conventional collinear AFMs using external fields. In this respect, light fields are particularly advantageous owing to their high tunability and non-contact nature. In fact, Floquet engineering, the tailoring of electronic and magnetic structures via time-periodic light, has emerged as a powerful tool.  For example, circularly polarized light (CPL) breaks the time-reversal (TR) symmetry,  leading to diverse phenomena such as the opening of surface state gap in topological insulators~\cite{wang2013observation}, photoinduced anomalous Hall effect~\cite{mciver2020light},  and so on~\cite{zhu2023floquet, zhou2023pseudospin, ning2024flexible, liu2018photoinduced, Zhang2021Anomalous, Grushin2014Floquet, Yan2016Tunable, hubener2017creating, bao2024manipulating, liu2025signatures, fan2025floquet,ghorashi2025dynamical}.

In this letter, we demonstrate through symmetry analysis that unconventional odd-parity magnetism can be induced in collinear AFMs by breaking a spin-preserving pseudo-TR symmetry while maintaining a spin-flipping inversion symmetry, which is achievable via CPL irradiation. Based on effective model analysis and Floquet theory, we reveal the emergence of novel Floquet odd-parity magnetic states, namely, Floquet $p$- and $f$-wave magnets in conventional AFM lattices under CPL. In addition, we find that antiferromagnetic Chern insulators with a high Chern number of two can be realized in the Floquet $f$-wave magnet. Notably, our first-principles calculations for CPL-driven MnPSe$_{3}$ directly confirm the Floquet-engineered odd-parity magnetic state. Our work not only establishes a novel class of odd-parity magnets but also paves the way for light-controlled spintronic functionalities.

\emph{Odd-parity magnetism induced from collinear AFM}. We start from conventional collinear AFMs where the two magnetic sublattices are related by inversion symmetry, namely, $PT$-symmetric AFM. We also focus on systems with negligible spin-orbit coupling (SOC), which can be described by spin group symmetries of the form $[S||R]$ ($S$ and $R$ operations act in the spin and real spaces, respectively)~\cite{jiang2024enumeration, chen2024enumeration, xiao2024spin, vsmejkal2022beyond}.  In such AFM systems,  two key symmetries can relate electronic states with opposite crystal momenta.  The first is  $[C_{2}||P]$, where  $C_2$ is the spin-space two-fold rotation about the axis perpendicular to the spin quantization direction and the inversion $P$ exchanges the two magnetic sublattices.  This symmetry enforces the energy dispersion relation $\varepsilon(s,\bm{k})=\varepsilon(-s,-\bm{k})$ ($s=\uparrow, \downarrow$). The second is $[C_{2}T||E]$, which consists of time reversal $T$ followed by a spin-space rotation $C_2$ ($E$ is the real-space identity operation) and holds in all collinear magnets~\cite{chen2024enumeration}. Under this symmetry, the electronic band structure satisfies $\varepsilon(s,\bm{k})=\varepsilon(s,-\bm{k})$ , as shown in Fig.~\ref{fig1}(a).

The combination of the above two spin-group symmetries is $PT$ symmetry and ensures the spin degeneracy in AFMs, as illustrated in Fig.~\ref{fig1}(b). Crucially, breaking the symmetry $[C_{2}T||E]$  while preserving $[C_{2}||P]$ lifts the spin-degeneracy, leaving $\varepsilon(s,\bm{k})=\varepsilon(-s,-\bm{k})$. As a result, the Fermi surface of the spin-split band structures should exhibit odd-parity symmetry, which generically takes a $p$-wave-like pattern, thus termed $p$-wave magnet, as shown in Fig.~\ref {fig1}(c). When an additional threefold rotation symmetry is respected in the $p$-wave magnet, as depicted in Fig.~\ref {fig1}(d), an $f$-wave magnet can emerge.  Remarkably, introducing CPL can satisfy the previous symmetry requirement, namely, breaking the spin-preserving pseudo-TR symmetry  $[C_{2}T||E]$  and preserving the spin-flipping inversion symmetry $[C_{2}||P]$ simultaneously. The latter arises because,  in SOC-free systems, CPL does not directly couple with spin and remains invariant under spatial inversion. Consequently,  Floquet engineering of AFMs with CPL provides a natural route to realizing odd-parity magnets, as demonstrated below.

\begin{figure}[!t]
	\centering
	\includegraphics[width=3.4in]{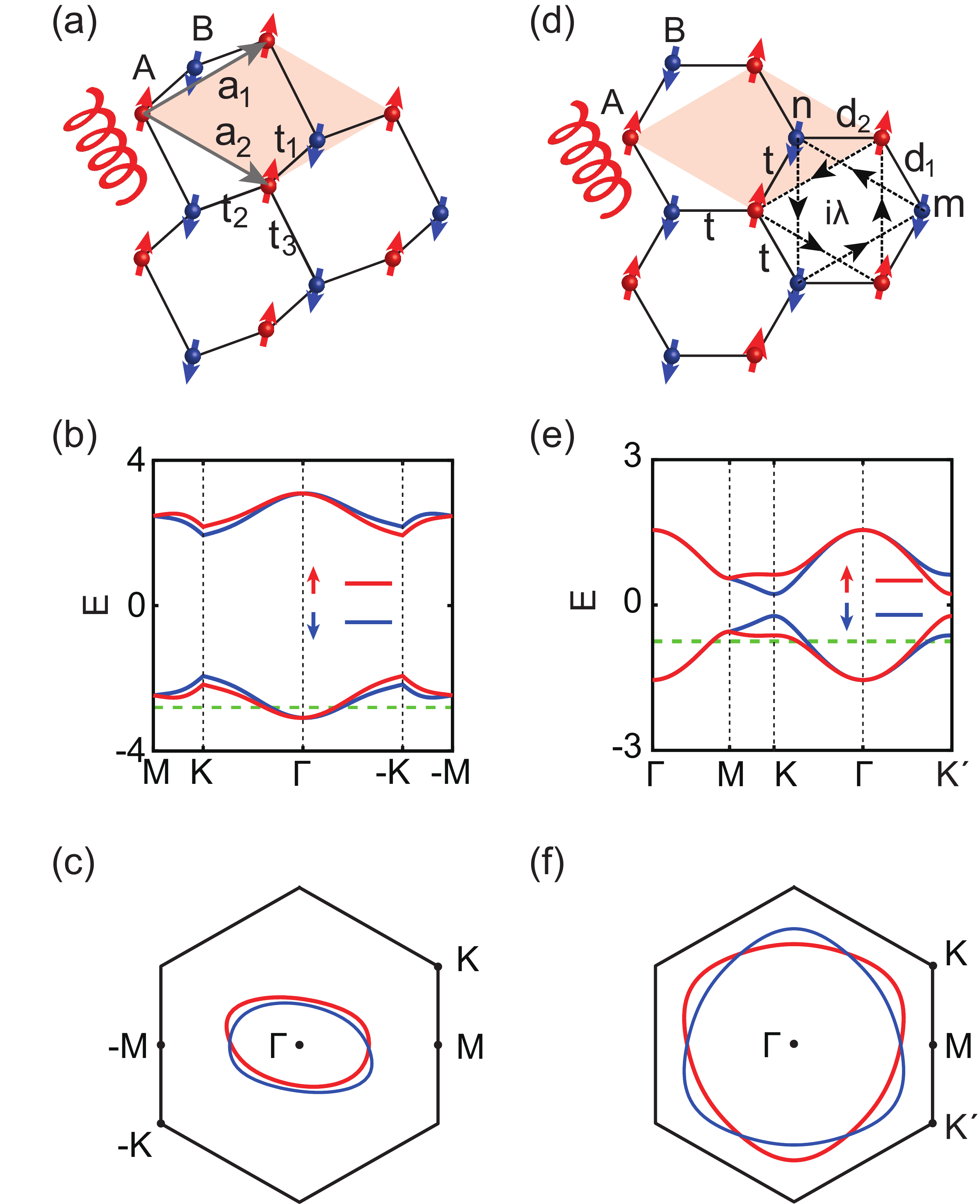}
	\caption{Lattice models of Floquet odd-parity collinear magnets. (a) The two-dimensional antiferromagnetic rhombic lattice with magnetic sublattices A  at $(0,0)$ and B  at $(0.5,0.8)$ in fractional coordinates. The parameters $t_1, t_2, t_3$ denote hoppings between A and its three nearest-neighbor B sites. The red spiral represents CPL illumination, which can induce effective hoppings (not explicitly shown). (b) Light-induced band structures calculated with dimensionless parameters $t_1=3, t_2=2, t_3=0.7,h_z=0.2, \tilde{A}a=2.6$ and $\hbar\omega=6$. (c) The spin-polarized  Fermi surfaces at  $E=-2.8$. (d) The two-dimensional antiferromagnetic honeycomb lattice with nearest-neighbor hopping  $t=1$. The light-induced effective next-nearest-neighbor hopping term with imaginary amplitude $i\lambda$ is explicitly shown. (e) Light-induced band structures with dimensionless parameters $ \tilde{A}a=2.6$ and $\hbar\omega=6$. (f) The $f$-wave symmetric spin-polarized Fermi surfaces at $ E=-0.75$. In (a) and (d), the shaded areas indicate the unit cells.} 
	\label{fig2}
\end{figure}

\emph{Floquet $p$- and $f$-wave magnets.} We consider a two-dimensional AFM model consisting of two magnetic sublattices A and B. To realize a  $p$-wave magnetic state, we design a low-symmetry structure, where sublattice B is positioned at a generic, non-high-symmetry point in the unit cell formed by the triangular A sublattices, as shown in Fig.~\ref{fig2}(a). The basis vectors are $\bm{a}_{1}=a(\sqrt{3}/{2},{1}/{2})$ and $\bm{a}_{2}=a(\sqrt{3}/{2},-{1}/{2})$, where $a$ is the lattice constant. The corresponding lattice Hamiltonian is given by
\begin{equation}
\begin{split}
H=\sum_{m,j}t_j c_{m}^\dagger c_{m+{\delta}_j}+ h_{z} \sum_{m} \chi_m c_{m}^\dagger s_z c_{m},
\end{split}
\label{eq2}
\end{equation}
where $t_{j}$ ($j=1,2,3$) denotes the hopping parameters between sublattice A and its three nearest-neighbor B sublattices (and vice versa) connected by displacement vectors $\delta_{j}$, the Pauli matrix $s_z$ acts in the spin subspaces, $h_z$ represents the staggered exchange field resulting from local magnetic moments with $\chi_m=+1(-1)$ for A (B) sublattice. At this stage, the band structure is doubly degenerate throughout the whole the Brillouin zone due to  $PT$ symmetry. 
 
Next, we introduce CPL into the system to break the spin degeneracy. In $\bm{k}$-space, the time-dependent Hamiltonian can be obtained by performing Peierls substitution $\bm{k}\rightarrow\bm{k}+e\bm{A}(t)/\hbar$,  where $\bm{A}(t)$ denotes the vector potential of the incident laser. Specifically,  the vector potential of CPL propagating along the $z$-axis can be expressed as $[A_x(t), A_y( t)]=A_0[\cos(\omega t),  \eta \sin(\omega t)]$. Where $\omega$ is the frequency of light, $A_0$ is the amplitude of the light,   and $\eta=\pm1$ corresponds to the right-handed CPL (RCPL) or left-handed CPL (LCPL), respectively. For notational simplicity, we define  $\tilde{A}={eA_0}/{\hbar}$, which has the same dimension as wavevector and will be used to represent the amplitude of light later. Using Floquet theory, we can map the time-periodic Hamiltonian onto an effective static Hamiltonian, whose spin-preserving TR symmetry is broken by CPL. In the off-resonant (high-frequency) regime, this effective  static Hamiltonian can be obtained by applying the off-resonant approximation~\cite{Kitagawa2011Transport,Mikami2016Brillouin,Marin2015Universal,Eckardt2015High}, and is given by 
\begin{equation}
\begin{split}
H_{\mbox{\tiny eff}}(\bm{k})&=H_0(\bm{k})+M(\bm{k}),\\
M(\bm{k})&=\sum_{n\ge 1}\frac{[H_{-n},H_{n}]}{n\omega}+O(\frac{1}{\omega^2}),
\end{split}
\label{high-frequency}
\end{equation}
where $H_n=\frac{1}{T}\int^T_0H(t)e^{in\omega t}dt$ is  the $n$-th Fourier component in the frequency space, and $T$ is the period of the light. Here,  $M(\bm{k})$ is primarily responsible for the light-induced modification of the band structures. We solve the Hamiltonian numerically and obtain the Floquet band structures shown in Fig.~\ref{fig2}(b), where the amplitude of RCPL $\tilde{A}a = 2.6$ is used. As expected, the bands are non-degenerate, and the spin-resolved Fermi surfaces at $E=-2.8$ shown in Fig.~\ref{fig2}(c) exhibit characteristics of a $p$-wave spin-splitting.

To realize an $f$-wave magnet, we consider a $C_{3}$-symmetric version of the above lattice structure, thereby restoring the characteristic honeycomb lattice structure in which the three nearest-neighbor hopping amplitudes become equivalent (denoted by $t$). For this structure, the form of $H_0(\bm{k})$ can be found in the Supplemental Material~\cite{SM}.  $M(\bm{k})$ up to the first order takes the form as 
\begin{equation}
\begin{split}
M(\bm{k})=4\eta\lambda (\cos{\frac{\sqrt{3}}{2}k_xa}-\cos{\frac{k_ya}{2}} ) \sin{\frac{k_ya}{2}}
\end{split}\sigma_z,
\end{equation}
where {$\lambda=-\sqrt{3}t^2 J_1(\tilde{A}a/\sqrt{3})^2/\omega$}, $\sigma{_z}$ is Pauli matrix acting on  A and B sublattices subspace and $J_1(\tilde{A}a/\sqrt{3})$ is the first-order Bessel function. 
%This term breaks $[C_{2}T||E]$ symmetry while preserving other rotational symmetry, which can be seen more clearly in real space. 
To see its contribution more clearly, we discretize $M(\bm{k})$ on a honeycomb lattice, as illustrated in Fig.~\ref{fig2}(d), and the corresponding effective Hamiltonian in real space takes the form:
\begin{equation}
\begin{split}
H_F=i\lambda \sum_{\langle\langle m,n\rangle \rangle}\nu_{mn}  c_{m}^{\dagger}c_{n},
\end{split}
\label{eq4}
\end{equation}
where $\langle\langle m,n\rangle \rangle$ represents the next-nearest-neighbors hopping, and   $\nu_{m n}$  is defined as $\nu_{m n}=\operatorname{sgn}\left[\left(\boldsymbol{d}_1 \times \boldsymbol{d}_2\right) \cdot \mathbf{z}\right]$ where $\bm{d}_{1,2}$ are the two bonds connecting next-nearest-neighboring sites. Notably, it is spin-independent and preserves the $[C_2||P]$ and $[E||C_3]$   symmetries of the original system. On the other hand, its imaginary hopping amplitude is odd under TR, thereby breaking the spin-preserving pseudo-TR symmetry $[C_2T||E]$.  Consequently, $f$-wave magnetism can be induced, which is confirmed by the calculated band structures and spin-polarized Fermi surfaces at $E=-2$ shown in Fig.~\ref{fig2}(e, f). The odd-parity spin splitting can be detected by a pump-probe experiment (see Supplemental Material for more details~\cite{SM}). We note that the imaginary hopping term in Eq.~(\ref{eq4}) was originally introduced in the spinless Haldane model through staggered magnetic fluxes.  Therefore, our model can also be viewed as a modified Haldane model with local magnetic moments.

\begin{figure}[!t]
	\centering
	\includegraphics[width=3.5in]{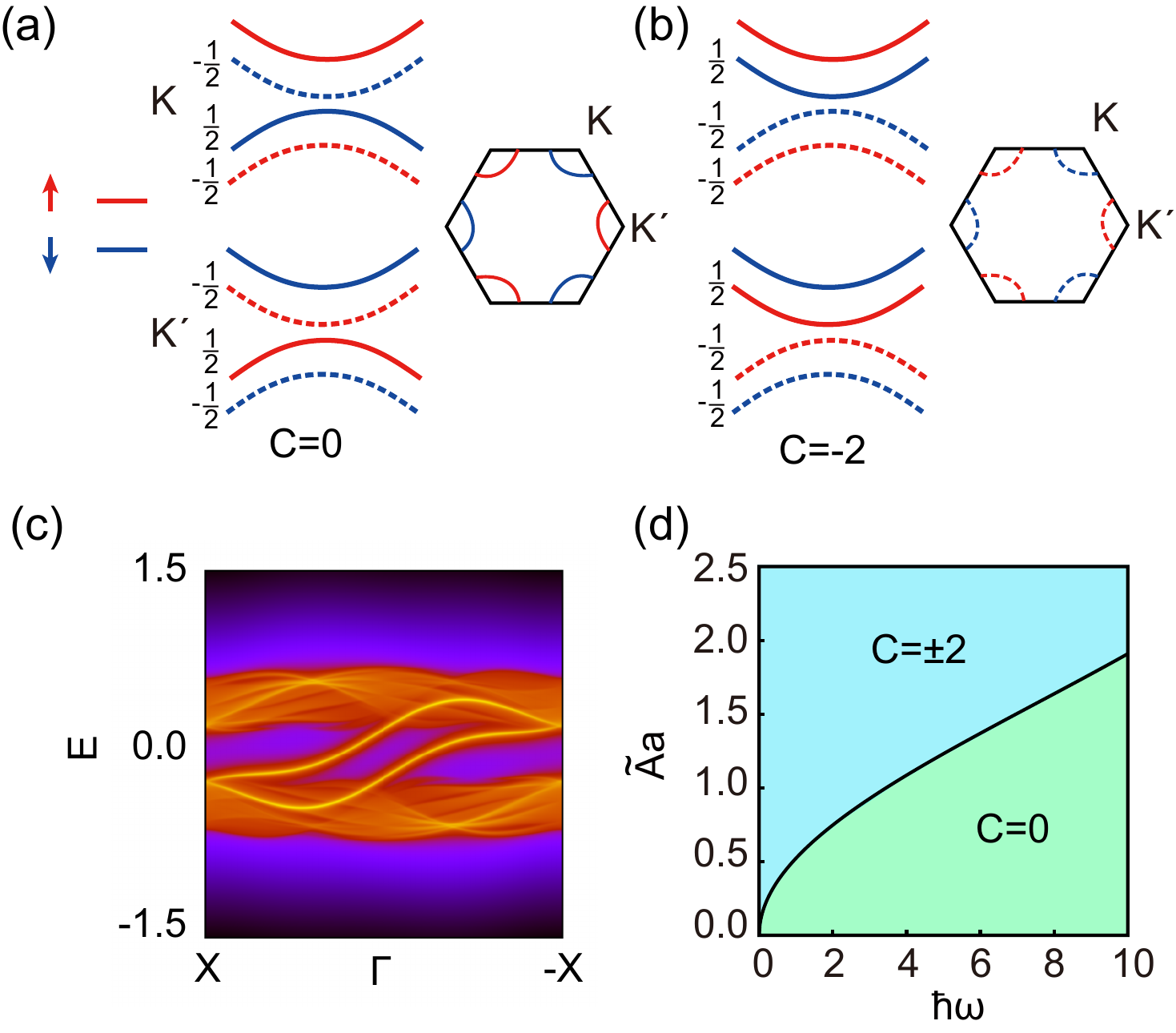}
	\caption{ Light-induced antiferromagnetic Chern states in the   $f$-wave magnet. (a,b) Schematic electronic bands at K and K' in the low- (a) and high-intensity (b) RCPL regimes. The dashed (solid) lines indicate bands with a Chern number of $\mathcal{C}=-1/2$ ($\mathcal{C}=1/2$). The red (blue) lines represent spin-up (spin-down) polarization. The spin-resolved Fermi surfaces are obtained by taking an energy cut below the valence band maximum.  (c) The surface states calculated for a light amplitude of $\tilde{A}a=1.73$ and $\hbar\omega=5$. (d) Topological phase diagram as a function of light amplitudes and frequency.}
	\label{fig3}
\end{figure}

\begin{figure*}[htbp]
	\centering
	\includegraphics[width=6.5in]{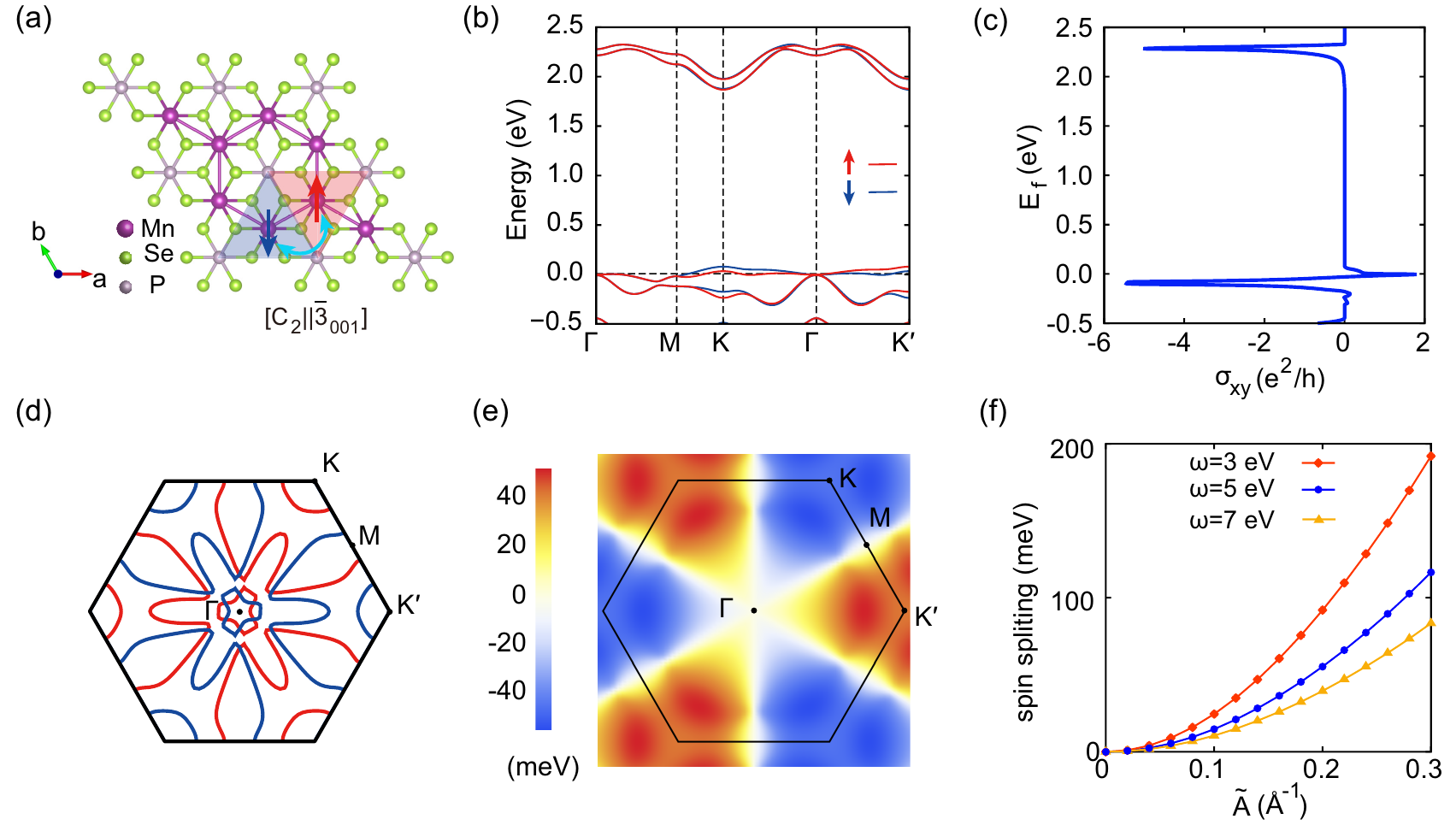}
	\caption{First-principles calculations of light-induced $f$-wave magnet in MnPSe$_{3}$. (a) Crystal structure of MnPSe$_{3}$. The magnetic atoms with opposite moments are related by the $[C_2||\overline{3}_{001}]$ symmetry, which is preserved under CPL irradiation. (b) Band structures under the irradiation of RCPL with an amplitude of $\tilde{A}=0.2$ \AA$^{-1}$ and a frequency at $\hbar\omega=5$ eV. (c) Anomalous Hall conductivity as a function of the Fermi level. (d) The spin-polarized Fermi surface at $E=0$ eV in (b). (e) Distribution of the spin splitting defined as the energy difference between the highest spin-up and spin-down valence bands.  (f) Variation of the maximal spin splitting as a function of light amplitude at different frequencies.}
	\label{fig4}
\end{figure*}

\emph{Antiferromagnetic Chern state in Floquet f-wave magnet.} In the above honeycomb-lattice-based Floquet $f$-wave magnet, the low-energy effective Hamiltonian up to linear order of momentum $\bm{q}$ around the $K$ and $K'$ points can be derived as
\begin{equation}
\begin{split}
H_{\text{eff}}(\bm{K}+\bm{q})&= v(q_x\sigma_y+q_y\sigma_x)+(s h_z- \eta \lambda')\sigma_z,\\
H_{\text{eff}}(\bm{K}'+\bm{q})&= v(-q_x\sigma_y+q_y\sigma_x)+(s h_z+\eta \lambda')\sigma_z,\\
\end{split}
\label{leKK}
\end{equation}
where $v=\sqrt{3}atJ_0(\tilde{A}a/\sqrt{3})/2$, {$\lambda'=3\sqrt{3}\lambda=-9 t^2J_1^2(\tilde{A}a/\sqrt{3})/\omega$}, $s=+1(-1)$ for $\uparrow$ ($\downarrow$), and $\eta=+1 (-1)$ for RCPL (LCPL) as mentioned earlier. The four-band Hamiltonian for each valley ($K/K'$) in Eq.~(\ref{leKK}) can be decomposed into a pair of spin-polarized massive Dirac cones, with their mass terms dependent on spin, valley, and, CPL degrees of freedom. 

In the case of RCPL, the corresponding Chern number of each spin-polarized Dirac cone (the integration of Berry curvatures over the lower Dirac band) is given by sign$(sh_z-\lambda')/2$ and sign$(sh_z+\lambda')/2$ at $K$ and $K'$ valleys, respectively. Consequently, in the low-intensity ($|\lambda'|<|h_z|$) RCPL regime, where the lower conduction band and the upper valence band at the $K$ ($K'$) valley stem from the spin-up (spin-down) polarized Dirac cone, as illustrated in Fig.~\ref{fig3}(a), the Chern numbers of the two valence bands are opposite, resulting in a vanishing total Chern number. When the light intensity increases to meet {$|\lambda'|=|h_z|$}, a simultaneous gap-closing and reopening process occurs for both the spin-up Dirac cone at $K$ and the spin-down Dirac cone at $K'$, each undergoing a topological phase transition accompanied by a Chern number change of $\mathcal{C}=-1$. This leads to a total Chern number of $\mathcal{C}=-2$ in the high-intensity ({$|\lambda'|>|h_z|$}) RCPL regime [see Fig.~\ref{fig3}(b)], which is validated by the emergence of two chiral edge states transversing the gap [Fig.~\ref{fig3}(c)]. Notably, the spin polarization of each band remains unchanged across the phase transition, and they all exhibit the $f$-wave feature of the Fermi surface. Similar analysis can be carried out for the RCPL case, yielding a Chern number of $\mathcal{C}=+2$ in the high-intensity regime. {It is noticed that the quantum anomalous Hall effect is induced by CPL and is independent of the spin degree of freedom}. In Fig.~\ref{fig3}(d), we present the full phase diagram as a function of the amplitude and frequency of CPL. { Besides the quantum anomalous Hall effect, this f-wave Chern state also presents a second-order pure spin current (see Supplemental Material for details~\cite{SM}). This effect can be intuitively understood from the combination of the Edelstein effect and Ohm's law, both of which scale linearly with the electric field.} In brief, the Floquet f-wave Chern state presented here unifies topology, odd-parity spin splitting, and nonlinear spin transport in a same material platform. A comparison with conventional Floquet Chern insulators\cite{Kitagawa2011Transport} is summarized in Supplemental Material ~\cite{SM}.

\emph{Material realization}. Following the above model calculation, we examine our design principle for Floquet odd-parity magnets in real materials,  based on first-principles calculations combined with Floquet theory~\cite{zhu2023floquet}. We take monolayer MnPSe${_3}$ as a prototype material, which is a Néel-type AFM in its ground state~\cite{Chittari2016Electronic}. As shown in Fig.~\ref {fig4}(a), the system exhibits the desired symmetry discussed earlier: the magnetic Mn atoms form a honeycomb lattice, and the neighboring Mn atoms carry opposite magnetic moments. Besides the spin-related inversion symmetry, the system also possesses a $[C_{2}||\overline{3}_{001}]$ symmetry that relates the two magnetic sublattices, where  $\overline{3}_{001}$ denotes a three-fold rotation about the z-axis followed by inversion. Thus, according to our previous theory, upon irradiation of CPL, monolayer MnPSe${_3}$ should exhibit $f$-wave spin splitting. This is evident by examining the band structures shown in Fig.~\ref{fig4}(b) and the spin-resolved Fermi surfaces shown in Fig.~\ref{fig4}(d), where RCPL with amplitude  $\tilde{A}=0.2$ \AA$^{-1}$ and frequency $\hbar \omega=5$ eV is employed in the calculations. We note that by applying uniaxial strain to break $\bar{3}_{001}$ symmetry, one can further induce $p$-wave spin-splitting via Floquet engineering (see Supplemental Material for more details~\cite{SM}).

Since $PT$ symmetry is broken upon irradiation of CPL, the anomalous Hall effect could emerge. We calculate the anomalous Hall conductivity for the irradiated MnPSe${_3}$ and present the result in Fig.~\ref{fig4}(c), which shows that the conductivity is sizable, reaching approximately $6e^2/h$ when the Fermi energy is tuned to around $-0.1$ eV. Lastly, we examine the spin splitting and its evolution as a function of laser amplitude and frequency. In Fig.~\ref{fig4}(e),  we present the calculated momentum-resolved spin splitting, which is defined as the energy difference between the highest spin-up and spin-down valence bands at each $\bm{k}$ point. Using the same parameters for CPL as before, the results show that the maximal spin splitting could reach 40 meV.  By varying the amplitude and frequency of the CPL, we find that the spin splitting can be widely tuned, as shown in Fig.~\ref{fig4}(f).

\emph{Conclusions}. In summary, we propose a new class of unconventional magnets with odd-parity spin splittings, referred to as odd-parity collinear magnets, which can be induced from conventional collinear AFMs by applying CPL. We demonstrate through an effective model and Floquet theory the existence of such light-induced odd-parity magnetic states, including Floquet $p$- and $f$-wave magnets in CPL-driven AFM lattices.  We also reveal a Floquet-engineered antiferromagnetic Chern insulator with $|\mathcal{C}| = 2$ in the $f$-wave magnet. Furthermore, first-principles calculations confirm the emergence of light-induced odd-parity magnetism in CPL-driven MnPSe$_3$, in excellent agreement with our model analysis. Our work not only expands the landscape of unconventional magnetism to include odd-parity collinear magnets but also highlights the significant role of Floquet engineering as a tuning knob of novel magnetic states and related quantum phenomena.

\emph{Note added}. Upon completion of this work, we became aware of two independent and concurrent studies by S. Huang et al. \cite{huang2025light} and B. Li et al. \cite{li2025floquet}, which report findings overlapping with some conclusions presented in this manuscript.

\section*{ACKNOWLEDGEMENTS}
This work is supported by the Natural Science Foundation of China (Grants No. 12074434, No.12534007, No.12504197, No.12504280),  the National Key Research and Development Program of China (Grants No. 2024YFA1409800), the Postdoctoral Fellowship Program of CPSF (Grant No. GZC20242012), Shandong Provincial Natural Science Foundation (Grant No. ZR2024QA095), the Fundamental Research Funds for the central Universities (Grant No. 23CX06063A), the 
Natural Science Foundation of Jiangsu Province (Grant No. BK20252117, No. BK20233001), the Youth Innovation Team Plan Project for the Higher Education Institution of Shandong Province  (Grant No. 2024KJN021), and the start-up grant from Eastern Institute of Technology, Ningbo.

\bibliography{ref}

\begin{thebibliography}{87}%
\makeatletter
\providecommand \@ifxundefined [1]{%
 \@ifx{#1\undefined}
}%
\providecommand \@ifnum [1]{%
 \ifnum #1\expandafter \@firstoftwo
 \else \expandafter \@secondoftwo
 \fi
}%
\providecommand \@ifx [1]{%
 \ifx #1\expandafter \@firstoftwo
 \else \expandafter \@secondoftwo
 \fi
}%
\providecommand \natexlab [1]{#1}%
\providecommand \enquote  [1]{``#1''}%
\providecommand \bibnamefont  [1]{#1}%
\providecommand \bibfnamefont [1]{#1}%
\providecommand \citenamefont [1]{#1}%
\providecommand \href@noop [0]{\@secondoftwo}%
\providecommand \href [0]{\begingroup \@sanitize@url \@href}%
\providecommand \@href[1]{\@@startlink{#1}\@@href}%
\providecommand \@@href[1]{\endgroup#1\@@endlink}%
\providecommand \@sanitize@url [0]{\catcode `\\12\catcode `\$12\catcode
  `\&12\catcode `\#12\catcode `\^12\catcode `\_12\catcode `\%12\relax}%
\providecommand \@@startlink[1]{}%
\providecommand \@@endlink[0]{}%
\providecommand \url  [0]{\begingroup\@sanitize@url \@url }%
\providecommand \@url [1]{\endgroup\@href {#1}{\urlprefix }}%
\providecommand \urlprefix  [0]{URL }%
\providecommand \Eprint [0]{\href }%
\providecommand \doibase [0]{https://doi.org/}%
\providecommand \selectlanguage [0]{\@gobble}%
\providecommand \bibinfo  [0]{\@secondoftwo}%
\providecommand \bibfield  [0]{\@secondoftwo}%
\providecommand \translation [1]{[#1]}%
\providecommand \BibitemOpen [0]{}%
\providecommand \bibitemStop [0]{}%
\providecommand \bibitemNoStop [0]{.\EOS\space}%
\providecommand \EOS [0]{\spacefactor3000\relax}%
\providecommand \BibitemShut  [1]{\csname bibitem#1\endcsname}%
\let\auto@bib@innerbib\@empty
%</preamble>
\bibitem [{\citenamefont {{\v{S}}mejkal}\ \emph
  {et~al.}(2022{\natexlab{a}})\citenamefont {{\v{S}}mejkal}, \citenamefont
  {Sinova},\ and\ \citenamefont {Jungwirth}}]{vsmejkal2022emerging}%
  \BibitemOpen
  \bibfield  {author} {\bibinfo {author} {\bibfnamefont {L.}~\bibnamefont
  {{\v{S}}mejkal}}, \bibinfo {author} {\bibfnamefont {J.}~\bibnamefont
  {Sinova}},\ and\ \bibinfo {author} {\bibfnamefont {T.}~\bibnamefont
  {Jungwirth}},\ }\bibfield  {title} {\bibinfo {title} {{Emerging research
  landscape of altermagnetism}},\ }\href
  {https://doi.org/10.1103/PhysRevX.12.040501} {\bibfield  {journal} {\bibinfo
  {journal} {Phys. Rev. X}\ }\textbf {\bibinfo {volume} {12}},\ \bibinfo
  {pages} {040501} (\bibinfo {year} {2022}{\natexlab{a}})}\BibitemShut
  {NoStop}%
\bibitem [{\citenamefont {Bai}\ \emph {et~al.}(2024)\citenamefont {Bai},
  \citenamefont {Feng}, \citenamefont {Liu}, \citenamefont {{\v{S}}mejkal},
  \citenamefont {Mokrousov},\ and\ \citenamefont
  {Yao}}]{bai2024altermagnetism}%
  \BibitemOpen
  \bibfield  {author} {\bibinfo {author} {\bibfnamefont {L.}~\bibnamefont
  {Bai}}, \bibinfo {author} {\bibfnamefont {W.}~\bibnamefont {Feng}}, \bibinfo
  {author} {\bibfnamefont {S.}~\bibnamefont {Liu}}, \bibinfo {author}
  {\bibfnamefont {L.}~\bibnamefont {{\v{S}}mejkal}}, \bibinfo {author}
  {\bibfnamefont {Y.}~\bibnamefont {Mokrousov}},\ and\ \bibinfo {author}
  {\bibfnamefont {Y.}~\bibnamefont {Yao}},\ }\bibfield  {title} {\bibinfo
  {title} {{{Altermagnetism: Exploring new frontiers in magnetism and
  spintronics}}},\ }\href {https://doi.org/10.1002/adfm.202409327} {\bibfield
  {journal} {\bibinfo  {journal} {Adv. Funct. Mater.}\ }\textbf {\bibinfo
  {volume} {34}},\ \bibinfo {pages} {2409327} (\bibinfo {year}
  {2024})}\BibitemShut {NoStop}%
\bibitem [{\citenamefont {Song}\ \emph
  {et~al.}(2025{\natexlab{a}})\citenamefont {Song}, \citenamefont {Bai},
  \citenamefont {Zhou}, \citenamefont {Han}, \citenamefont {Reichlova},
  \citenamefont {Dil}, \citenamefont {Liu}, \citenamefont {Chen},\ and\
  \citenamefont {Pan}}]{song2025altermagnets}%
  \BibitemOpen
  \bibfield  {author} {\bibinfo {author} {\bibfnamefont {C.}~\bibnamefont
  {Song}}, \bibinfo {author} {\bibfnamefont {H.}~\bibnamefont {Bai}}, \bibinfo
  {author} {\bibfnamefont {Z.}~\bibnamefont {Zhou}}, \bibinfo {author}
  {\bibfnamefont {L.}~\bibnamefont {Han}}, \bibinfo {author} {\bibfnamefont
  {H.}~\bibnamefont {Reichlova}}, \bibinfo {author} {\bibfnamefont {J.~H.}\
  \bibnamefont {Dil}}, \bibinfo {author} {\bibfnamefont {J.}~\bibnamefont
  {Liu}}, \bibinfo {author} {\bibfnamefont {X.}~\bibnamefont {Chen}},\ and\
  \bibinfo {author} {\bibfnamefont {F.}~\bibnamefont {Pan}},\ }\bibfield
  {title} {\bibinfo {title} {{Altermagnets as a new class of functional
  materials}},\ }\href {https://doi.org/10.1038/s41578-025-00779-1} {\bibfield
  {journal} {\bibinfo  {journal} {Nat. Rev. Mater.}\ }\textbf {\bibinfo
  {volume} {10}},\ \bibinfo {pages} {473} (\bibinfo {year}
  {2025}{\natexlab{a}})}\BibitemShut {NoStop}%
\bibitem [{\citenamefont {{\v{S}}mejkal}\ \emph
  {et~al.}(2022{\natexlab{b}})\citenamefont {{\v{S}}mejkal}, \citenamefont
  {Sinova},\ and\ \citenamefont {Jungwirth}}]{vsmejkal2022beyond}%
  \BibitemOpen
  \bibfield  {author} {\bibinfo {author} {\bibfnamefont {L.}~\bibnamefont
  {{\v{S}}mejkal}}, \bibinfo {author} {\bibfnamefont {J.}~\bibnamefont
  {Sinova}},\ and\ \bibinfo {author} {\bibfnamefont {T.}~\bibnamefont
  {Jungwirth}},\ }\bibfield  {title} {\bibinfo {title} {{Beyond conventional
  ferromagnetism and antiferromagnetism: A phase with nonrelativistic spin and
  crystal rotation symmetry}},\ }\href
  {https://doi.org/10.1103/PhysRevX.12.031042} {\bibfield  {journal} {\bibinfo
  {journal} {Phys. Rev. X}\ }\textbf {\bibinfo {volume} {12}},\ \bibinfo
  {pages} {031042} (\bibinfo {year} {2022}{\natexlab{b}})}\BibitemShut
  {NoStop}%
\bibitem [{\citenamefont {Wu}\ \emph {et~al.}(2007)\citenamefont {Wu},
  \citenamefont {Sun}, \citenamefont {Fradkin},\ and\ \citenamefont
  {Zhang}}]{wu2007fermi}%
  \BibitemOpen
  \bibfield  {author} {\bibinfo {author} {\bibfnamefont {C.}~\bibnamefont
  {Wu}}, \bibinfo {author} {\bibfnamefont {K.}~\bibnamefont {Sun}}, \bibinfo
  {author} {\bibfnamefont {E.}~\bibnamefont {Fradkin}},\ and\ \bibinfo {author}
  {\bibfnamefont {S.-C.}\ \bibnamefont {Zhang}},\ }\bibfield  {title} {\bibinfo
  {title} {{Fermi liquid instabilities in the spin channel}},\ }\href
  {https://doi.org/10.1103/PhysRevB.75.115103} {\bibfield  {journal} {\bibinfo
  {journal} {Phys. Rev. B}\ }\textbf {\bibinfo {volume} {75}},\ \bibinfo
  {pages} {115103} (\bibinfo {year} {2007})}\BibitemShut {NoStop}%
\bibitem [{\citenamefont {Ma}\ \emph {et~al.}(2021)\citenamefont {Ma},
  \citenamefont {Hu}, \citenamefont {Li}, \citenamefont {Liu}, \citenamefont
  {Yao}, \citenamefont {Jia},\ and\ \citenamefont
  {Liu}}]{ma2021multifunctional}%
  \BibitemOpen
  \bibfield  {author} {\bibinfo {author} {\bibfnamefont {H.-Y.}\ \bibnamefont
  {Ma}}, \bibinfo {author} {\bibfnamefont {M.}~\bibnamefont {Hu}}, \bibinfo
  {author} {\bibfnamefont {N.}~\bibnamefont {Li}}, \bibinfo {author}
  {\bibfnamefont {J.}~\bibnamefont {Liu}}, \bibinfo {author} {\bibfnamefont
  {W.}~\bibnamefont {Yao}}, \bibinfo {author} {\bibfnamefont {J.-F.}\
  \bibnamefont {Jia}},\ and\ \bibinfo {author} {\bibfnamefont {J.}~\bibnamefont
  {Liu}},\ }\bibfield  {title} {\bibinfo {title} {{Multifunctional
  antiferromagnetic materials with giant piezomagnetism and noncollinear spin
  current}},\ }\href {https://doi.org/10.1038/s41467-021-23127-7} {\bibfield
  {journal} {\bibinfo  {journal} {Nat. Commun.}\ }\textbf {\bibinfo {volume}
  {12}},\ \bibinfo {pages} {2846} (\bibinfo {year} {2021})}\BibitemShut
  {NoStop}%
\bibitem [{\citenamefont {Mazin}\ \emph {et~al.}(2023)\citenamefont {Mazin},
  \citenamefont {Gonz{\'a}lez-Hern{\'a}ndez},\ and\ \citenamefont
  {{\v{S}}mejkal}}]{mazin2023induced}%
  \BibitemOpen
  \bibfield  {author} {\bibinfo {author} {\bibfnamefont {I.}~\bibnamefont
  {Mazin}}, \bibinfo {author} {\bibfnamefont {R.}~\bibnamefont
  {Gonz{\'a}lez-Hern{\'a}ndez}},\ and\ \bibinfo {author} {\bibfnamefont
  {L.}~\bibnamefont {{\v{S}}mejkal}},\ }\bibfield  {title} {\bibinfo {title}
  {{Induced Monolayer Altermagnetism in MnP(S,Se)$_3 $ and FeSe}},\ }\href@noop
  {} {\bibfield  {journal} {\bibinfo  {journal} {arXiv preprint
  arXiv:2309.02355}\ } (\bibinfo {year} {2023})}\BibitemShut {NoStop}%
\bibitem [{\citenamefont {Zeng}\ and\ \citenamefont
  {Zhao}(2024)}]{zeng2024description}%
  \BibitemOpen
  \bibfield  {author} {\bibinfo {author} {\bibfnamefont {S.}~\bibnamefont
  {Zeng}}\ and\ \bibinfo {author} {\bibfnamefont {Y.-J.}\ \bibnamefont
  {Zhao}},\ }\bibfield  {title} {\bibinfo {title} {{Description of
  two-dimensional altermagnetism: Categorization using spin group theory}},\
  }\href {https://doi.org/10.1103/PhysRevB.110.054406} {\bibfield  {journal}
  {\bibinfo  {journal} {Phys. Rev. B}\ }\textbf {\bibinfo {volume} {110}},\
  \bibinfo {pages} {054406} (\bibinfo {year} {2024})}\BibitemShut {NoStop}%
\bibitem [{\citenamefont {Regmi}\ \emph {et~al.}(2025)\citenamefont {Regmi},
  \citenamefont {Bhandari}, \citenamefont {Thapa}, \citenamefont {Hao},
  \citenamefont {Sharma}, \citenamefont {McKenzie}, \citenamefont {Chen},
  \citenamefont {Nayak}, \citenamefont {El~Gazzah}, \citenamefont {M{\'a}rkus}
  \emph {et~al.}}]{regmi2025altermagnetism}%
  \BibitemOpen
  \bibfield  {author} {\bibinfo {author} {\bibfnamefont {R.~B.}\ \bibnamefont
  {Regmi}}, \bibinfo {author} {\bibfnamefont {H.}~\bibnamefont {Bhandari}},
  \bibinfo {author} {\bibfnamefont {B.}~\bibnamefont {Thapa}}, \bibinfo
  {author} {\bibfnamefont {Y.}~\bibnamefont {Hao}}, \bibinfo {author}
  {\bibfnamefont {N.}~\bibnamefont {Sharma}}, \bibinfo {author} {\bibfnamefont
  {J.}~\bibnamefont {McKenzie}}, \bibinfo {author} {\bibfnamefont
  {X.}~\bibnamefont {Chen}}, \bibinfo {author} {\bibfnamefont {A.}~\bibnamefont
  {Nayak}}, \bibinfo {author} {\bibfnamefont {M.}~\bibnamefont {El~Gazzah}},
  \bibinfo {author} {\bibfnamefont {B.~G.}\ \bibnamefont {M{\'a}rkus}}, \emph
  {et~al.},\ }\bibfield  {title} {\bibinfo {title} {{Altermagnetism in the
  layered intercalated transition metal dichalcogenide CoNb$_4$Se$_8$}},\
  }\href {https://doi.org/10.1038/s41467-025-58642-4} {\bibfield  {journal}
  {\bibinfo  {journal} {Nat. Commun.}\ }\textbf {\bibinfo {volume} {16}},\
  \bibinfo {pages} {4399} (\bibinfo {year} {2025})}\BibitemShut {NoStop}%
\bibitem [{\citenamefont {Wang}\ \emph {et~al.}(2024)\citenamefont {Wang},
  \citenamefont {Wang}, \citenamefont {Liu}, \citenamefont {Zhang},\ and\
  \citenamefont {Zhang}}]{wang2024electric}%
  \BibitemOpen
  \bibfield  {author} {\bibinfo {author} {\bibfnamefont {D.}~\bibnamefont
  {Wang}}, \bibinfo {author} {\bibfnamefont {H.}~\bibnamefont {Wang}}, \bibinfo
  {author} {\bibfnamefont {L.}~\bibnamefont {Liu}}, \bibinfo {author}
  {\bibfnamefont {J.}~\bibnamefont {Zhang}},\ and\ \bibinfo {author}
  {\bibfnamefont {H.}~\bibnamefont {Zhang}},\ }\bibfield  {title} {\bibinfo
  {title} {{Electric-field-induced switchable two-dimensional altermagnets}},\
  }\href {https://doi.org/10.1021/acs.nanolett.4c05384} {\bibfield  {journal}
  {\bibinfo  {journal} {Nano Lett.}\ }\textbf {\bibinfo {volume} {25}},\
  \bibinfo {pages} {498} (\bibinfo {year} {2024})}\BibitemShut {NoStop}%
\bibitem [{\citenamefont {Liu}\ \emph {et~al.}(2024)\citenamefont {Liu},
  \citenamefont {Yu},\ and\ \citenamefont {Liu}}]{Liu2024Twisted}%
  \BibitemOpen
  \bibfield  {author} {\bibinfo {author} {\bibfnamefont {Y.}~\bibnamefont
  {Liu}}, \bibinfo {author} {\bibfnamefont {J.}~\bibnamefont {Yu}},\ and\
  \bibinfo {author} {\bibfnamefont {C.-C.}\ \bibnamefont {Liu}},\ }\bibfield
  {title} {\bibinfo {title} {{Twisted Magnetic Van der Waals Bilayers: An Ideal
  Platform for Altermagnetism}},\ }\href
  {https://doi.org/10.1103/PhysRevLett.133.206702} {\bibfield  {journal}
  {\bibinfo  {journal} {Phys. Rev. Lett.}\ }\textbf {\bibinfo {volume} {133}},\
  \bibinfo {pages} {206702} (\bibinfo {year} {2024})}\BibitemShut {NoStop}%
\bibitem [{\citenamefont {Duan}\ \emph {et~al.}(2025)\citenamefont {Duan},
  \citenamefont {Zhang}, \citenamefont {Zhu}, \citenamefont {Liu},
  \citenamefont {Zhang}, \citenamefont {{\v{Z}}uti{\'c}},\ and\ \citenamefont
  {Zhou}}]{duan2025antiferroelectric}%
  \BibitemOpen
  \bibfield  {author} {\bibinfo {author} {\bibfnamefont {X.}~\bibnamefont
  {Duan}}, \bibinfo {author} {\bibfnamefont {J.}~\bibnamefont {Zhang}},
  \bibinfo {author} {\bibfnamefont {Z.}~\bibnamefont {Zhu}}, \bibinfo {author}
  {\bibfnamefont {Y.}~\bibnamefont {Liu}}, \bibinfo {author} {\bibfnamefont
  {Z.}~\bibnamefont {Zhang}}, \bibinfo {author} {\bibfnamefont
  {I.}~\bibnamefont {{\v{Z}}uti{\'c}}},\ and\ \bibinfo {author} {\bibfnamefont
  {T.}~\bibnamefont {Zhou}},\ }\bibfield  {title} {\bibinfo {title}
  {{Antiferroelectric altermagnets: Antiferroelectricity alters magnets}},\
  }\href {https://doi.org/10.1103/PhysRevLett.134.106801} {\bibfield  {journal}
  {\bibinfo  {journal} {Phys. Rev. Lett.}\ }\textbf {\bibinfo {volume} {134}},\
  \bibinfo {pages} {106801} (\bibinfo {year} {2025})}\BibitemShut {NoStop}%
\bibitem [{\citenamefont {Leeb}\ \emph {et~al.}(2024)\citenamefont {Leeb},
  \citenamefont {Mook}, \citenamefont {\ifmmode~\check{S}\else
  \v{S}\fi{}mejkal},\ and\ \citenamefont {Knolle}}]{Leeb2024Spontaneous}%
  \BibitemOpen
  \bibfield  {author} {\bibinfo {author} {\bibfnamefont {V.}~\bibnamefont
  {Leeb}}, \bibinfo {author} {\bibfnamefont {A.}~\bibnamefont {Mook}}, \bibinfo
  {author} {\bibfnamefont {L.}~\bibnamefont {\ifmmode~\check{S}\else
  \v{S}\fi{}mejkal}},\ and\ \bibinfo {author} {\bibfnamefont {J.}~\bibnamefont
  {Knolle}},\ }\bibfield  {title} {\bibinfo {title} {{Spontaneous Formation of
  Altermagnetism from Orbital Ordering}},\ }\href
  {https://doi.org/10.1103/PhysRevLett.132.236701} {\bibfield  {journal}
  {\bibinfo  {journal} {Phys. Rev. Lett.}\ }\textbf {\bibinfo {volume} {132}},\
  \bibinfo {pages} {236701} (\bibinfo {year} {2024})}\BibitemShut {NoStop}%
\bibitem [{\citenamefont {Zhu}\ \emph {et~al.}(2025)\citenamefont {Zhu},
  \citenamefont {Huo}, \citenamefont {Feng}, \citenamefont {Zhang},
  \citenamefont {Yang},\ and\ \citenamefont {Guo}}]{Zhu2025Design}%
  \BibitemOpen
  \bibfield  {author} {\bibinfo {author} {\bibfnamefont {X.}~\bibnamefont
  {Zhu}}, \bibinfo {author} {\bibfnamefont {X.}~\bibnamefont {Huo}}, \bibinfo
  {author} {\bibfnamefont {S.}~\bibnamefont {Feng}}, \bibinfo {author}
  {\bibfnamefont {S.-B.}\ \bibnamefont {Zhang}}, \bibinfo {author}
  {\bibfnamefont {S.~A.}\ \bibnamefont {Yang}},\ and\ \bibinfo {author}
  {\bibfnamefont {H.}~\bibnamefont {Guo}},\ }\bibfield  {title} {\bibinfo
  {title} {{Design of Altermagnetic Models from Spin Clusters}},\ }\href
  {https://doi.org/10.1103/PhysRevLett.134.166701} {\bibfield  {journal}
  {\bibinfo  {journal} {Phys. Rev. Lett.}\ }\textbf {\bibinfo {volume} {134}},\
  \bibinfo {pages} {166701} (\bibinfo {year} {2025})}\BibitemShut {NoStop}%
\bibitem [{\citenamefont {Gu}\ \emph {et~al.}(2025)\citenamefont {Gu},
  \citenamefont {Liu}, \citenamefont {Zhu}, \citenamefont {Yananose},
  \citenamefont {Chen}, \citenamefont {Hu}, \citenamefont {Stroppa},\ and\
  \citenamefont {Liu}}]{Gu2025Ferroelectric}%
  \BibitemOpen
  \bibfield  {author} {\bibinfo {author} {\bibfnamefont {M.}~\bibnamefont
  {Gu}}, \bibinfo {author} {\bibfnamefont {Y.}~\bibnamefont {Liu}}, \bibinfo
  {author} {\bibfnamefont {H.}~\bibnamefont {Zhu}}, \bibinfo {author}
  {\bibfnamefont {K.}~\bibnamefont {Yananose}}, \bibinfo {author}
  {\bibfnamefont {X.}~\bibnamefont {Chen}}, \bibinfo {author} {\bibfnamefont
  {Y.}~\bibnamefont {Hu}}, \bibinfo {author} {\bibfnamefont {A.}~\bibnamefont
  {Stroppa}},\ and\ \bibinfo {author} {\bibfnamefont {Q.}~\bibnamefont {Liu}},\
  }\bibfield  {title} {\bibinfo {title} {{Ferroelectric Switchable
  Altermagnetism}},\ }\href {https://doi.org/10.1103/PhysRevLett.134.106802}
  {\bibfield  {journal} {\bibinfo  {journal} {Phys. Rev. Lett.}\ }\textbf
  {\bibinfo {volume} {134}},\ \bibinfo {pages} {106802} (\bibinfo {year}
  {2025})}\BibitemShut {NoStop}%
\bibitem [{\citenamefont {Reimers}\ \emph {et~al.}(2024)\citenamefont
  {Reimers}, \citenamefont {Odenbreit}, \citenamefont {{\v{S}}mejkal},
  \citenamefont {Strocov}, \citenamefont {Constantinou}, \citenamefont
  {Hellenes}, \citenamefont {Jaeschke~Ubiergo}, \citenamefont {Campos},
  \citenamefont {Bharadwaj}, \citenamefont {Chakraborty} \emph
  {et~al.}}]{reimers2024direct}%
  \BibitemOpen
  \bibfield  {author} {\bibinfo {author} {\bibfnamefont {S.}~\bibnamefont
  {Reimers}}, \bibinfo {author} {\bibfnamefont {L.}~\bibnamefont {Odenbreit}},
  \bibinfo {author} {\bibfnamefont {L.}~\bibnamefont {{\v{S}}mejkal}}, \bibinfo
  {author} {\bibfnamefont {V.~N.}\ \bibnamefont {Strocov}}, \bibinfo {author}
  {\bibfnamefont {P.}~\bibnamefont {Constantinou}}, \bibinfo {author}
  {\bibfnamefont {A.~B.}\ \bibnamefont {Hellenes}}, \bibinfo {author}
  {\bibfnamefont {R.}~\bibnamefont {Jaeschke~Ubiergo}}, \bibinfo {author}
  {\bibfnamefont {W.~H.}\ \bibnamefont {Campos}}, \bibinfo {author}
  {\bibfnamefont {V.~K.}\ \bibnamefont {Bharadwaj}}, \bibinfo {author}
  {\bibfnamefont {A.}~\bibnamefont {Chakraborty}}, \emph {et~al.},\ }\bibfield
  {title} {\bibinfo {title} {{Direct observation of altermagnetic band
  splitting in CrSb thin films}},\ }\href
  {https://doi.org/10.1038/s41467-024-46476-5} {\bibfield  {journal} {\bibinfo
  {journal} {Nat. Commun.}\ }\textbf {\bibinfo {volume} {15}},\ \bibinfo
  {pages} {2116} (\bibinfo {year} {2024})}\BibitemShut {NoStop}%
\bibitem [{\citenamefont {Zhou}\ \emph {et~al.}(2025)\citenamefont {Zhou},
  \citenamefont {Cheng}, \citenamefont {Hu}, \citenamefont {Chu}, \citenamefont
  {Bai}, \citenamefont {Han}, \citenamefont {Liu}, \citenamefont {Pan},\ and\
  \citenamefont {Song}}]{zhou2025manipulation}%
  \BibitemOpen
  \bibfield  {author} {\bibinfo {author} {\bibfnamefont {Z.}~\bibnamefont
  {Zhou}}, \bibinfo {author} {\bibfnamefont {X.}~\bibnamefont {Cheng}},
  \bibinfo {author} {\bibfnamefont {M.}~\bibnamefont {Hu}}, \bibinfo {author}
  {\bibfnamefont {R.}~\bibnamefont {Chu}}, \bibinfo {author} {\bibfnamefont
  {H.}~\bibnamefont {Bai}}, \bibinfo {author} {\bibfnamefont {L.}~\bibnamefont
  {Han}}, \bibinfo {author} {\bibfnamefont {J.}~\bibnamefont {Liu}}, \bibinfo
  {author} {\bibfnamefont {F.}~\bibnamefont {Pan}},\ and\ \bibinfo {author}
  {\bibfnamefont {C.}~\bibnamefont {Song}},\ }\bibfield  {title} {\bibinfo
  {title} {{Manipulation of the altermagnetic order in CrSb via crystal
  symmetry}},\ }\href {https://doi.org/10.1038/s41586-024-08436-3} {\bibfield
  {journal} {\bibinfo  {journal} {Nature}\ }\textbf {\bibinfo {volume} {638}},\
  \bibinfo {pages} {645} (\bibinfo {year} {2025})}\BibitemShut {NoStop}%
\bibitem [{\citenamefont {Karube}\ \emph {et~al.}(2022)\citenamefont {Karube},
  \citenamefont {Tanaka}, \citenamefont {Sugawara}, \citenamefont {Kadoguchi},
  \citenamefont {Kohda},\ and\ \citenamefont {Nitta}}]{Karube2022Observation}%
  \BibitemOpen
  \bibfield  {author} {\bibinfo {author} {\bibfnamefont {S.}~\bibnamefont
  {Karube}}, \bibinfo {author} {\bibfnamefont {T.}~\bibnamefont {Tanaka}},
  \bibinfo {author} {\bibfnamefont {D.}~\bibnamefont {Sugawara}}, \bibinfo
  {author} {\bibfnamefont {N.}~\bibnamefont {Kadoguchi}}, \bibinfo {author}
  {\bibfnamefont {M.}~\bibnamefont {Kohda}},\ and\ \bibinfo {author}
  {\bibfnamefont {J.}~\bibnamefont {Nitta}},\ }\bibfield  {title} {\bibinfo
  {title} {{Observation of Spin-Splitter Torque in Collinear Antiferromagnetic
  ${\mathrm{RuO}}_{2}$}},\ }\href
  {https://doi.org/10.1103/PhysRevLett.129.137201} {\bibfield  {journal}
  {\bibinfo  {journal} {Phys. Rev. Lett.}\ }\textbf {\bibinfo {volume} {129}},\
  \bibinfo {pages} {137201} (\bibinfo {year} {2022})}\BibitemShut {NoStop}%
\bibitem [{\citenamefont {Bose}\ \emph {et~al.}(2022)\citenamefont {Bose},
  \citenamefont {Schreiber}, \citenamefont {Jain}, \citenamefont {Shao},
  \citenamefont {Nair}, \citenamefont {Sun}, \citenamefont {Zhang},
  \citenamefont {Muller}, \citenamefont {Tsymbal}, \citenamefont {Schlom} \emph
  {et~al.}}]{bose2022tilted}%
  \BibitemOpen
  \bibfield  {author} {\bibinfo {author} {\bibfnamefont {A.}~\bibnamefont
  {Bose}}, \bibinfo {author} {\bibfnamefont {N.~J.}\ \bibnamefont {Schreiber}},
  \bibinfo {author} {\bibfnamefont {R.}~\bibnamefont {Jain}}, \bibinfo {author}
  {\bibfnamefont {D.-F.}\ \bibnamefont {Shao}}, \bibinfo {author}
  {\bibfnamefont {H.~P.}\ \bibnamefont {Nair}}, \bibinfo {author}
  {\bibfnamefont {J.}~\bibnamefont {Sun}}, \bibinfo {author} {\bibfnamefont
  {X.~S.}\ \bibnamefont {Zhang}}, \bibinfo {author} {\bibfnamefont {D.~A.}\
  \bibnamefont {Muller}}, \bibinfo {author} {\bibfnamefont {E.~Y.}\
  \bibnamefont {Tsymbal}}, \bibinfo {author} {\bibfnamefont {D.~G.}\
  \bibnamefont {Schlom}}, \emph {et~al.},\ }\bibfield  {title} {\bibinfo
  {title} {{Tilted spin current generated by the collinear antiferromagnet
  ruthenium dioxide}},\ }\href {https://doi.org/10.1038/s41928-022-00744-8}
  {\bibfield  {journal} {\bibinfo  {journal} {Nat. Electron.}\ }\textbf
  {\bibinfo {volume} {5}},\ \bibinfo {pages} {267} (\bibinfo {year}
  {2022})}\BibitemShut {NoStop}%
\bibitem [{\citenamefont {Fedchenko}\ \emph {et~al.}(2024)\citenamefont
  {Fedchenko}, \citenamefont {Min{\'a}r}, \citenamefont {Akashdeep},
  \citenamefont {D’Souza}, \citenamefont {Vasilyev}, \citenamefont {Tkach},
  \citenamefont {Odenbreit}, \citenamefont {Nguyen}, \citenamefont
  {Kutnyakhov}, \citenamefont {Wind} \emph
  {et~al.}}]{fedchenko2024observation}%
  \BibitemOpen
  \bibfield  {author} {\bibinfo {author} {\bibfnamefont {O.}~\bibnamefont
  {Fedchenko}}, \bibinfo {author} {\bibfnamefont {J.}~\bibnamefont
  {Min{\'a}r}}, \bibinfo {author} {\bibfnamefont {A.}~\bibnamefont
  {Akashdeep}}, \bibinfo {author} {\bibfnamefont {S.~W.}\ \bibnamefont
  {D’Souza}}, \bibinfo {author} {\bibfnamefont {D.}~\bibnamefont {Vasilyev}},
  \bibinfo {author} {\bibfnamefont {O.}~\bibnamefont {Tkach}}, \bibinfo
  {author} {\bibfnamefont {L.}~\bibnamefont {Odenbreit}}, \bibinfo {author}
  {\bibfnamefont {Q.}~\bibnamefont {Nguyen}}, \bibinfo {author} {\bibfnamefont
  {D.}~\bibnamefont {Kutnyakhov}}, \bibinfo {author} {\bibfnamefont
  {N.}~\bibnamefont {Wind}}, \emph {et~al.},\ }\bibfield  {title} {\bibinfo
  {title} {{Observation of time-reversal symmetry breaking in the band
  structure of altermagnetic RuO$_2$}},\ }\href
  {https://doi.org/10.1126/sciadv.adj4883} {\bibfield  {journal} {\bibinfo
  {journal} {Sci. Adv.}\ }\textbf {\bibinfo {volume} {10}},\ \bibinfo {pages}
  {eadj4883} (\bibinfo {year} {2024})}\BibitemShut {NoStop}%
\bibitem [{\citenamefont {Ding}\ \emph {et~al.}(2024)\citenamefont {Ding},
  \citenamefont {Jiang}, \citenamefont {Chen}, \citenamefont {Tao},
  \citenamefont {Liu}, \citenamefont {Li}, \citenamefont {Liu}, \citenamefont
  {Sun}, \citenamefont {Cheng}, \citenamefont {Liu}, \citenamefont {Yang},
  \citenamefont {Zhang}, \citenamefont {Deng}, \citenamefont {Jing},
  \citenamefont {Huang}, \citenamefont {Shi}, \citenamefont {Ye}, \citenamefont
  {Qiao}, \citenamefont {Wang}, \citenamefont {Guo}, \citenamefont {Feng},\
  and\ \citenamefont {Shen}}]{Ding2024Large}%
  \BibitemOpen
  \bibfield  {author} {\bibinfo {author} {\bibfnamefont {J.}~\bibnamefont
  {Ding}}, \bibinfo {author} {\bibfnamefont {Z.}~\bibnamefont {Jiang}},
  \bibinfo {author} {\bibfnamefont {X.}~\bibnamefont {Chen}}, \bibinfo {author}
  {\bibfnamefont {Z.}~\bibnamefont {Tao}}, \bibinfo {author} {\bibfnamefont
  {Z.}~\bibnamefont {Liu}}, \bibinfo {author} {\bibfnamefont {T.}~\bibnamefont
  {Li}}, \bibinfo {author} {\bibfnamefont {J.}~\bibnamefont {Liu}}, \bibinfo
  {author} {\bibfnamefont {J.}~\bibnamefont {Sun}}, \bibinfo {author}
  {\bibfnamefont {J.}~\bibnamefont {Cheng}}, \bibinfo {author} {\bibfnamefont
  {J.}~\bibnamefont {Liu}}, \bibinfo {author} {\bibfnamefont {Y.}~\bibnamefont
  {Yang}}, \bibinfo {author} {\bibfnamefont {R.}~\bibnamefont {Zhang}},
  \bibinfo {author} {\bibfnamefont {L.}~\bibnamefont {Deng}}, \bibinfo {author}
  {\bibfnamefont {W.}~\bibnamefont {Jing}}, \bibinfo {author} {\bibfnamefont
  {Y.}~\bibnamefont {Huang}}, \bibinfo {author} {\bibfnamefont
  {Y.}~\bibnamefont {Shi}}, \bibinfo {author} {\bibfnamefont {M.}~\bibnamefont
  {Ye}}, \bibinfo {author} {\bibfnamefont {S.}~\bibnamefont {Qiao}}, \bibinfo
  {author} {\bibfnamefont {Y.}~\bibnamefont {Wang}}, \bibinfo {author}
  {\bibfnamefont {Y.}~\bibnamefont {Guo}}, \bibinfo {author} {\bibfnamefont
  {D.}~\bibnamefont {Feng}},\ and\ \bibinfo {author} {\bibfnamefont
  {D.}~\bibnamefont {Shen}},\ }\bibfield  {title} {\bibinfo {title} {{Large
  Band Splitting in $g$-Wave Altermagnet CrSb}},\ }\href
  {https://doi.org/10.1103/PhysRevLett.133.206401} {\bibfield  {journal}
  {\bibinfo  {journal} {Phys. Rev. Lett.}\ }\textbf {\bibinfo {volume} {133}},\
  \bibinfo {pages} {206401} (\bibinfo {year} {2024})}\BibitemShut {NoStop}%
\bibitem [{\citenamefont {Han}\ \emph {et~al.}(2024)\citenamefont {Han},
  \citenamefont {Fu}, \citenamefont {Peng}, \citenamefont {Cheng},
  \citenamefont {Dai}, \citenamefont {Liu}, \citenamefont {Li}, \citenamefont
  {Zhang}, \citenamefont {Zhu}, \citenamefont {Bai} \emph
  {et~al.}}]{han2024electrical}%
  \BibitemOpen
  \bibfield  {author} {\bibinfo {author} {\bibfnamefont {L.}~\bibnamefont
  {Han}}, \bibinfo {author} {\bibfnamefont {X.}~\bibnamefont {Fu}}, \bibinfo
  {author} {\bibfnamefont {R.}~\bibnamefont {Peng}}, \bibinfo {author}
  {\bibfnamefont {X.}~\bibnamefont {Cheng}}, \bibinfo {author} {\bibfnamefont
  {J.}~\bibnamefont {Dai}}, \bibinfo {author} {\bibfnamefont {L.}~\bibnamefont
  {Liu}}, \bibinfo {author} {\bibfnamefont {Y.}~\bibnamefont {Li}}, \bibinfo
  {author} {\bibfnamefont {Y.}~\bibnamefont {Zhang}}, \bibinfo {author}
  {\bibfnamefont {W.}~\bibnamefont {Zhu}}, \bibinfo {author} {\bibfnamefont
  {H.}~\bibnamefont {Bai}}, \emph {et~al.},\ }\bibfield  {title} {\bibinfo
  {title} {{Electrical 180 switching of N{\'e}el vector in spin-splitting
  antiferromagnet}},\ }\href {https://doi.org/10.1126/sciadv.adn0479}
  {\bibfield  {journal} {\bibinfo  {journal} {Sci. Adv.}\ }\textbf {\bibinfo
  {volume} {10}},\ \bibinfo {pages} {eadn0479} (\bibinfo {year}
  {2024})}\BibitemShut {NoStop}%
\bibitem [{\citenamefont {Reichlova}\ \emph {et~al.}(2024)\citenamefont
  {Reichlova}, \citenamefont {Lopes~Seeger}, \citenamefont
  {Gonz{\'a}lez-Hern{\'a}ndez}, \citenamefont {Kounta}, \citenamefont
  {Schlitz}, \citenamefont {Kriegner}, \citenamefont {Ritzinger}, \citenamefont
  {Lammel}, \citenamefont {Leivisk{\"a}}, \citenamefont {Birk~Hellenes} \emph
  {et~al.}}]{reichlova2024observation}%
  \BibitemOpen
  \bibfield  {author} {\bibinfo {author} {\bibfnamefont {H.}~\bibnamefont
  {Reichlova}}, \bibinfo {author} {\bibfnamefont {R.}~\bibnamefont
  {Lopes~Seeger}}, \bibinfo {author} {\bibfnamefont {R.}~\bibnamefont
  {Gonz{\'a}lez-Hern{\'a}ndez}}, \bibinfo {author} {\bibfnamefont
  {I.}~\bibnamefont {Kounta}}, \bibinfo {author} {\bibfnamefont
  {R.}~\bibnamefont {Schlitz}}, \bibinfo {author} {\bibfnamefont
  {D.}~\bibnamefont {Kriegner}}, \bibinfo {author} {\bibfnamefont
  {P.}~\bibnamefont {Ritzinger}}, \bibinfo {author} {\bibfnamefont
  {M.}~\bibnamefont {Lammel}}, \bibinfo {author} {\bibfnamefont
  {M.}~\bibnamefont {Leivisk{\"a}}}, \bibinfo {author} {\bibfnamefont
  {A.}~\bibnamefont {Birk~Hellenes}}, \emph {et~al.},\ }\bibfield  {title}
  {\bibinfo {title} {{Observation of a spontaneous anomalous Hall response in
  the Mn$_5$Si$_3$ d-wave altermagnet candidate}},\ }\href
  {https://doi.org/10.1038/s41467-024-48493-w} {\bibfield  {journal} {\bibinfo
  {journal} {Nat. Commun.}\ }\textbf {\bibinfo {volume} {15}},\ \bibinfo
  {pages} {4961} (\bibinfo {year} {2024})}\BibitemShut {NoStop}%
\bibitem [{\citenamefont {Lee}\ \emph {et~al.}(2024)\citenamefont {Lee},
  \citenamefont {Lee}, \citenamefont {Jung}, \citenamefont {Jung},
  \citenamefont {Kim}, \citenamefont {Lee}, \citenamefont {Seok}, \citenamefont
  {Kim}, \citenamefont {Park}, \citenamefont {{\v{S}}mejkal} \emph
  {et~al.}}]{lee2024broken}%
  \BibitemOpen
  \bibfield  {author} {\bibinfo {author} {\bibfnamefont {S.}~\bibnamefont
  {Lee}}, \bibinfo {author} {\bibfnamefont {S.}~\bibnamefont {Lee}}, \bibinfo
  {author} {\bibfnamefont {S.}~\bibnamefont {Jung}}, \bibinfo {author}
  {\bibfnamefont {J.}~\bibnamefont {Jung}}, \bibinfo {author} {\bibfnamefont
  {D.}~\bibnamefont {Kim}}, \bibinfo {author} {\bibfnamefont {Y.}~\bibnamefont
  {Lee}}, \bibinfo {author} {\bibfnamefont {B.}~\bibnamefont {Seok}}, \bibinfo
  {author} {\bibfnamefont {J.}~\bibnamefont {Kim}}, \bibinfo {author}
  {\bibfnamefont {B.~G.}\ \bibnamefont {Park}}, \bibinfo {author}
  {\bibfnamefont {L.}~\bibnamefont {{\v{S}}mejkal}}, \emph {et~al.},\
  }\bibfield  {title} {\bibinfo {title} {{Broken Kramers degeneracy in
  altermagnetic MnTe}},\ }\href
  {https://doi.org/10.1103/PhysRevLett.132.036702} {\bibfield  {journal}
  {\bibinfo  {journal} {Phys. Rev. Lett.}\ }\textbf {\bibinfo {volume} {132}},\
  \bibinfo {pages} {036702} (\bibinfo {year} {2024})}\BibitemShut {NoStop}%
\bibitem [{\citenamefont {Osumi}\ \emph {et~al.}(2024)\citenamefont {Osumi},
  \citenamefont {Souma}, \citenamefont {Aoyama}, \citenamefont {Yamauchi},
  \citenamefont {Honma}, \citenamefont {Nakayama}, \citenamefont {Takahashi},
  \citenamefont {Ohgushi},\ and\ \citenamefont {Sato}}]{Osumi2024Observation}%
  \BibitemOpen
  \bibfield  {author} {\bibinfo {author} {\bibfnamefont {T.}~\bibnamefont
  {Osumi}}, \bibinfo {author} {\bibfnamefont {S.}~\bibnamefont {Souma}},
  \bibinfo {author} {\bibfnamefont {T.}~\bibnamefont {Aoyama}}, \bibinfo
  {author} {\bibfnamefont {K.}~\bibnamefont {Yamauchi}}, \bibinfo {author}
  {\bibfnamefont {A.}~\bibnamefont {Honma}}, \bibinfo {author} {\bibfnamefont
  {K.}~\bibnamefont {Nakayama}}, \bibinfo {author} {\bibfnamefont
  {T.}~\bibnamefont {Takahashi}}, \bibinfo {author} {\bibfnamefont
  {K.}~\bibnamefont {Ohgushi}},\ and\ \bibinfo {author} {\bibfnamefont
  {T.}~\bibnamefont {Sato}},\ }\bibfield  {title} {\bibinfo {title}
  {{Observation of a giant band splitting in altermagnetic MnTe}},\ }\href
  {https://doi.org/10.1103/PhysRevB.109.115102} {\bibfield  {journal} {\bibinfo
   {journal} {Phys. Rev. B}\ }\textbf {\bibinfo {volume} {109}},\ \bibinfo
  {pages} {115102} (\bibinfo {year} {2024})}\BibitemShut {NoStop}%
\bibitem [{\citenamefont {Krempask{\`y}}\ \emph {et~al.}(2024)\citenamefont
  {Krempask{\`y}}, \citenamefont {{\v{S}}mejkal}, \citenamefont {D’souza},
  \citenamefont {Hajlaoui}, \citenamefont {Springholz}, \citenamefont
  {Uhl{\'\i}{\v{r}}ov{\'a}}, \citenamefont {Alarab}, \citenamefont
  {Constantinou}, \citenamefont {Strocov}, \citenamefont {Usanov} \emph
  {et~al.}}]{krempasky2024altermagnetic}%
  \BibitemOpen
  \bibfield  {author} {\bibinfo {author} {\bibfnamefont {J.}~\bibnamefont
  {Krempask{\`y}}}, \bibinfo {author} {\bibfnamefont {L.}~\bibnamefont
  {{\v{S}}mejkal}}, \bibinfo {author} {\bibfnamefont {S.}~\bibnamefont
  {D’souza}}, \bibinfo {author} {\bibfnamefont {M.}~\bibnamefont {Hajlaoui}},
  \bibinfo {author} {\bibfnamefont {G.}~\bibnamefont {Springholz}}, \bibinfo
  {author} {\bibfnamefont {K.}~\bibnamefont {Uhl{\'\i}{\v{r}}ov{\'a}}},
  \bibinfo {author} {\bibfnamefont {F.}~\bibnamefont {Alarab}}, \bibinfo
  {author} {\bibfnamefont {P.}~\bibnamefont {Constantinou}}, \bibinfo {author}
  {\bibfnamefont {V.}~\bibnamefont {Strocov}}, \bibinfo {author} {\bibfnamefont
  {D.}~\bibnamefont {Usanov}}, \emph {et~al.},\ }\bibfield  {title} {\bibinfo
  {title} {{Altermagnetic lifting of Kramers spin degeneracy}},\ }\href
  {https://doi.org/10.1038/s41586-023-06907-7} {\bibfield  {journal} {\bibinfo
  {journal} {Nature}\ }\textbf {\bibinfo {volume} {626}},\ \bibinfo {pages}
  {517} (\bibinfo {year} {2024})}\BibitemShut {NoStop}%
\bibitem [{\citenamefont {Mazin}(2023)}]{Mazin2023altermagnetism}%
  \BibitemOpen
  \bibfield  {author} {\bibinfo {author} {\bibfnamefont {I.~I.}\ \bibnamefont
  {Mazin}},\ }\bibfield  {title} {\bibinfo {title} {{Altermagnetism in MnTe:
  Origin, predicted manifestations, and routes to detwinning}},\ }\href
  {https://doi.org/10.1103/PhysRevB.107.L100418} {\bibfield  {journal}
  {\bibinfo  {journal} {Phys. Rev. B}\ }\textbf {\bibinfo {volume} {107}},\
  \bibinfo {pages} {L100418} (\bibinfo {year} {2023})}\BibitemShut {NoStop}%
\bibitem [{\citenamefont {Hariki}\ \emph {et~al.}(2024)\citenamefont {Hariki},
  \citenamefont {Dal~Din}, \citenamefont {Amin}, \citenamefont {Yamaguchi},
  \citenamefont {Badura}, \citenamefont {Kriegner}, \citenamefont {Edmonds},
  \citenamefont {Campion}, \citenamefont {Wadley}, \citenamefont {Backes},
  \citenamefont {Veiga}, \citenamefont {Dhesi}, \citenamefont {Springholz},
  \citenamefont {\ifmmode~\check{S}\else \v{S}\fi{}mejkal}, \citenamefont
  {V\'yborn\'y}, \citenamefont {Jungwirth},\ and\ \citenamefont
  {Kune\ifmmode~\check{s}\else \v{s}\fi{}}}]{Hariki2024X-Ray}%
  \BibitemOpen
  \bibfield  {author} {\bibinfo {author} {\bibfnamefont {A.}~\bibnamefont
  {Hariki}}, \bibinfo {author} {\bibfnamefont {A.}~\bibnamefont {Dal~Din}},
  \bibinfo {author} {\bibfnamefont {O.~J.}\ \bibnamefont {Amin}}, \bibinfo
  {author} {\bibfnamefont {T.}~\bibnamefont {Yamaguchi}}, \bibinfo {author}
  {\bibfnamefont {A.}~\bibnamefont {Badura}}, \bibinfo {author} {\bibfnamefont
  {D.}~\bibnamefont {Kriegner}}, \bibinfo {author} {\bibfnamefont {K.~W.}\
  \bibnamefont {Edmonds}}, \bibinfo {author} {\bibfnamefont {R.~P.}\
  \bibnamefont {Campion}}, \bibinfo {author} {\bibfnamefont {P.}~\bibnamefont
  {Wadley}}, \bibinfo {author} {\bibfnamefont {D.}~\bibnamefont {Backes}},
  \bibinfo {author} {\bibfnamefont {L.~S.~I.}\ \bibnamefont {Veiga}}, \bibinfo
  {author} {\bibfnamefont {S.~S.}\ \bibnamefont {Dhesi}}, \bibinfo {author}
  {\bibfnamefont {G.}~\bibnamefont {Springholz}}, \bibinfo {author}
  {\bibfnamefont {L.}~\bibnamefont {\ifmmode~\check{S}\else \v{S}\fi{}mejkal}},
  \bibinfo {author} {\bibfnamefont {K.}~\bibnamefont {V\'yborn\'y}}, \bibinfo
  {author} {\bibfnamefont {T.}~\bibnamefont {Jungwirth}},\ and\ \bibinfo
  {author} {\bibfnamefont {J.}~\bibnamefont {Kune\ifmmode~\check{s}\else
  \v{s}\fi{}}},\ }\bibfield  {title} {\bibinfo {title} {{X-Ray Magnetic
  Circular Dichroism in Altermagnetic $\ensuremath{\alpha}$-MnTe}},\ }\href
  {https://doi.org/10.1103/PhysRevLett.132.176701} {\bibfield  {journal}
  {\bibinfo  {journal} {Phys. Rev. Lett.}\ }\textbf {\bibinfo {volume} {132}},\
  \bibinfo {pages} {176701} (\bibinfo {year} {2024})}\BibitemShut {NoStop}%
\bibitem [{\citenamefont {Jiang}\ \emph {et~al.}(2025)\citenamefont {Jiang},
  \citenamefont {Hu}, \citenamefont {Bai}, \citenamefont {Song}, \citenamefont
  {Mu}, \citenamefont {Qu}, \citenamefont {Li}, \citenamefont {Zhu},
  \citenamefont {Pi}, \citenamefont {Wei} \emph {et~al.}}]{jiang2025metallic}%
  \BibitemOpen
  \bibfield  {author} {\bibinfo {author} {\bibfnamefont {B.}~\bibnamefont
  {Jiang}}, \bibinfo {author} {\bibfnamefont {M.}~\bibnamefont {Hu}}, \bibinfo
  {author} {\bibfnamefont {J.}~\bibnamefont {Bai}}, \bibinfo {author}
  {\bibfnamefont {Z.}~\bibnamefont {Song}}, \bibinfo {author} {\bibfnamefont
  {C.}~\bibnamefont {Mu}}, \bibinfo {author} {\bibfnamefont {G.}~\bibnamefont
  {Qu}}, \bibinfo {author} {\bibfnamefont {W.}~\bibnamefont {Li}}, \bibinfo
  {author} {\bibfnamefont {W.}~\bibnamefont {Zhu}}, \bibinfo {author}
  {\bibfnamefont {H.}~\bibnamefont {Pi}}, \bibinfo {author} {\bibfnamefont
  {Z.}~\bibnamefont {Wei}}, \emph {et~al.},\ }\bibfield  {title} {\bibinfo
  {title} {{A metallic room-temperature d-wave altermagnet}},\ }\href
  {https://doi.org/10.1038/s41567-025-02822-y} {\bibfield  {journal} {\bibinfo
  {journal} {Nat. Phys.}\ }\textbf {\bibinfo {volume} {21}},\ \bibinfo {pages}
  {754} (\bibinfo {year} {2025})}\BibitemShut {NoStop}%
\bibitem [{\citenamefont {Zhang}\ \emph {et~al.}(2025)\citenamefont {Zhang},
  \citenamefont {Cheng}, \citenamefont {Yin}, \citenamefont {Liu},
  \citenamefont {Deng}, \citenamefont {Qiao}, \citenamefont {Shi},
  \citenamefont {Zhang}, \citenamefont {Lin}, \citenamefont {Liu} \emph
  {et~al.}}]{zhang2025crystal}%
  \BibitemOpen
  \bibfield  {author} {\bibinfo {author} {\bibfnamefont {F.}~\bibnamefont
  {Zhang}}, \bibinfo {author} {\bibfnamefont {X.}~\bibnamefont {Cheng}},
  \bibinfo {author} {\bibfnamefont {Z.}~\bibnamefont {Yin}}, \bibinfo {author}
  {\bibfnamefont {C.}~\bibnamefont {Liu}}, \bibinfo {author} {\bibfnamefont
  {L.}~\bibnamefont {Deng}}, \bibinfo {author} {\bibfnamefont {Y.}~\bibnamefont
  {Qiao}}, \bibinfo {author} {\bibfnamefont {Z.}~\bibnamefont {Shi}}, \bibinfo
  {author} {\bibfnamefont {S.}~\bibnamefont {Zhang}}, \bibinfo {author}
  {\bibfnamefont {J.}~\bibnamefont {Lin}}, \bibinfo {author} {\bibfnamefont
  {Z.}~\bibnamefont {Liu}}, \emph {et~al.},\ }\bibfield  {title} {\bibinfo
  {title} {{Crystal-symmetry-paired spin--valley locking in a layered
  room-temperature metallic altermagnet candidate}},\ }\href
  {https://doi.org/10.1038/s41567-025-02864-2} {\bibfield  {journal} {\bibinfo
  {journal} {Nat. Phys.}\ }\textbf {\bibinfo {volume} {21}},\ \bibinfo {pages}
  {760} (\bibinfo {year} {2025})}\BibitemShut {NoStop}%
\bibitem [{\citenamefont {Banerjee}\ and\ \citenamefont
  {Scheurer}(2024)}]{Banerjee2024Altermagnetic}%
  \BibitemOpen
  \bibfield  {author} {\bibinfo {author} {\bibfnamefont {S.}~\bibnamefont
  {Banerjee}}\ and\ \bibinfo {author} {\bibfnamefont {M.~S.}\ \bibnamefont
  {Scheurer}},\ }\bibfield  {title} {\bibinfo {title} {{Altermagnetic
  superconducting diode effect}},\ }\href
  {https://doi.org/10.1103/PhysRevB.110.024503} {\bibfield  {journal} {\bibinfo
   {journal} {Phys. Rev. B}\ }\textbf {\bibinfo {volume} {110}},\ \bibinfo
  {pages} {024503} (\bibinfo {year} {2024})}\BibitemShut {NoStop}%
\bibitem [{\citenamefont {{\v{S}}mejkal}\ \emph {et~al.}(2023)\citenamefont
  {{\v{S}}mejkal}, \citenamefont {Marmodoro}, \citenamefont {Ahn},
  \citenamefont {Gonz{\'a}lez-Hern{\'a}ndez}, \citenamefont {Turek},
  \citenamefont {Mankovsky}, \citenamefont {Ebert}, \citenamefont {D’Souza},
  \citenamefont {{\v{S}}ipr}, \citenamefont {Sinova} \emph
  {et~al.}}]{vsmejkal2023chiral}%
  \BibitemOpen
  \bibfield  {author} {\bibinfo {author} {\bibfnamefont {L.}~\bibnamefont
  {{\v{S}}mejkal}}, \bibinfo {author} {\bibfnamefont {A.}~\bibnamefont
  {Marmodoro}}, \bibinfo {author} {\bibfnamefont {K.-H.}\ \bibnamefont {Ahn}},
  \bibinfo {author} {\bibfnamefont {R.}~\bibnamefont
  {Gonz{\'a}lez-Hern{\'a}ndez}}, \bibinfo {author} {\bibfnamefont
  {I.}~\bibnamefont {Turek}}, \bibinfo {author} {\bibfnamefont
  {S.}~\bibnamefont {Mankovsky}}, \bibinfo {author} {\bibfnamefont
  {H.}~\bibnamefont {Ebert}}, \bibinfo {author} {\bibfnamefont {S.~W.}\
  \bibnamefont {D’Souza}}, \bibinfo {author} {\bibfnamefont {O.}~\bibnamefont
  {{\v{S}}ipr}}, \bibinfo {author} {\bibfnamefont {J.}~\bibnamefont {Sinova}},
  \emph {et~al.},\ }\bibfield  {title} {\bibinfo {title} {{Chiral magnons in
  altermagnetic RuO$_2$}},\ }\href
  {https://doi.org/10.1103/PhysRevLett.131.256703} {\bibfield  {journal}
  {\bibinfo  {journal} {Phys. Rev. Lett.}\ }\textbf {\bibinfo {volume} {131}},\
  \bibinfo {pages} {256703} (\bibinfo {year} {2023})}\BibitemShut {NoStop}%
\bibitem [{\citenamefont {Xu}\ \emph {et~al.}(2025)\citenamefont {Xu},
  \citenamefont {Wu}, \citenamefont {Zhi}, \citenamefont {Cao}, \citenamefont
  {Dai}, \citenamefont {Cao}, \citenamefont {Wang},\ and\ \citenamefont
  {Lin}}]{xu2025altermagnetic}%
  \BibitemOpen
  \bibfield  {author} {\bibinfo {author} {\bibfnamefont {C.}~\bibnamefont
  {Xu}}, \bibinfo {author} {\bibfnamefont {S.}~\bibnamefont {Wu}}, \bibinfo
  {author} {\bibfnamefont {G.-X.}\ \bibnamefont {Zhi}}, \bibinfo {author}
  {\bibfnamefont {G.}~\bibnamefont {Cao}}, \bibinfo {author} {\bibfnamefont
  {J.}~\bibnamefont {Dai}}, \bibinfo {author} {\bibfnamefont {C.}~\bibnamefont
  {Cao}}, \bibinfo {author} {\bibfnamefont {X.}~\bibnamefont {Wang}},\ and\
  \bibinfo {author} {\bibfnamefont {H.-Q.}\ \bibnamefont {Lin}},\ }\bibfield
  {title} {\bibinfo {title} {{Altermagnetic ground state in distorted Kagome
  metal CsCr$_3$Sb$_5$}},\ }\href {https://doi.org/10.1038/s41467-025-58446-6}
  {\bibfield  {journal} {\bibinfo  {journal} {Nat. Commun.}\ }\textbf {\bibinfo
  {volume} {16}},\ \bibinfo {pages} {3114} (\bibinfo {year}
  {2025})}\BibitemShut {NoStop}%
\bibitem [{\citenamefont {Che}\ \emph {et~al.}(2025)\citenamefont {Che},
  \citenamefont {Lv}, \citenamefont {Wu},\ and\ \citenamefont
  {Yang}}]{Che2025Engineering}%
  \BibitemOpen
  \bibfield  {author} {\bibinfo {author} {\bibfnamefont {Y.}~\bibnamefont
  {Che}}, \bibinfo {author} {\bibfnamefont {H.}~\bibnamefont {Lv}}, \bibinfo
  {author} {\bibfnamefont {X.}~\bibnamefont {Wu}},\ and\ \bibinfo {author}
  {\bibfnamefont {J.}~\bibnamefont {Yang}},\ }\bibfield  {title} {\bibinfo
  {title} {{Engineering Altermagnetic States in Two-Dimensional Square
  Tessellations}},\ }\href {https://doi.org/10.1103/v38b-5by1} {\bibfield
  {journal} {\bibinfo  {journal} {Phys. Rev. Lett.}\ }\textbf {\bibinfo
  {volume} {135}},\ \bibinfo {pages} {036701} (\bibinfo {year}
  {2025})}\BibitemShut {NoStop}%
\bibitem [{\citenamefont {Gao}\ \emph {et~al.}(2025)\citenamefont {Gao},
  \citenamefont {Qu}, \citenamefont {Zeng}, \citenamefont {Liu}, \citenamefont
  {Wen}, \citenamefont {Sun}, \citenamefont {Guo},\ and\ \citenamefont
  {Lu}}]{gao2025ai}%
  \BibitemOpen
  \bibfield  {author} {\bibinfo {author} {\bibfnamefont {Z.-F.}\ \bibnamefont
  {Gao}}, \bibinfo {author} {\bibfnamefont {S.}~\bibnamefont {Qu}}, \bibinfo
  {author} {\bibfnamefont {B.}~\bibnamefont {Zeng}}, \bibinfo {author}
  {\bibfnamefont {Y.}~\bibnamefont {Liu}}, \bibinfo {author} {\bibfnamefont
  {J.-R.}\ \bibnamefont {Wen}}, \bibinfo {author} {\bibfnamefont
  {H.}~\bibnamefont {Sun}}, \bibinfo {author} {\bibfnamefont {P.-J.}\
  \bibnamefont {Guo}},\ and\ \bibinfo {author} {\bibfnamefont {Z.-Y.}\
  \bibnamefont {Lu}},\ }\bibfield  {title} {\bibinfo {title} {{AI-accelerated
  discovery of altermagnetic materials}},\ }\href
  {https://doi.org/10.1093/nsr/nwaf066} {\bibfield  {journal} {\bibinfo
  {journal} {Natl. Sci. Rev.}\ }\textbf {\bibinfo {volume} {12}},\ \bibinfo
  {pages} {nwaf066} (\bibinfo {year} {2025})}\BibitemShut {NoStop}%
\bibitem [{\citenamefont {Pan}\ \emph {et~al.}(2024)\citenamefont {Pan},
  \citenamefont {Zhou}, \citenamefont {Lyu}, \citenamefont {Xiao},
  \citenamefont {Yang},\ and\ \citenamefont {Sun}}]{Pan2024General}%
  \BibitemOpen
  \bibfield  {author} {\bibinfo {author} {\bibfnamefont {B.}~\bibnamefont
  {Pan}}, \bibinfo {author} {\bibfnamefont {P.}~\bibnamefont {Zhou}}, \bibinfo
  {author} {\bibfnamefont {P.}~\bibnamefont {Lyu}}, \bibinfo {author}
  {\bibfnamefont {H.}~\bibnamefont {Xiao}}, \bibinfo {author} {\bibfnamefont
  {X.}~\bibnamefont {Yang}},\ and\ \bibinfo {author} {\bibfnamefont
  {L.}~\bibnamefont {Sun}},\ }\bibfield  {title} {\bibinfo {title} {{General
  Stacking Theory for Altermagnetism in Bilayer Systems}},\ }\href
  {https://doi.org/10.1103/PhysRevLett.133.166701} {\bibfield  {journal}
  {\bibinfo  {journal} {Phys. Rev. Lett.}\ }\textbf {\bibinfo {volume} {133}},\
  \bibinfo {pages} {166701} (\bibinfo {year} {2024})}\BibitemShut {NoStop}%
\bibitem [{\citenamefont {Lin}\ \emph {et~al.}(2025)\citenamefont {Lin},
  \citenamefont {Zhang}, \citenamefont {Lu},\ and\ \citenamefont
  {Xie}}]{Lin2025Coulomb}%
  \BibitemOpen
  \bibfield  {author} {\bibinfo {author} {\bibfnamefont {H.-J.}\ \bibnamefont
  {Lin}}, \bibinfo {author} {\bibfnamefont {S.-B.}\ \bibnamefont {Zhang}},
  \bibinfo {author} {\bibfnamefont {H.-Z.}\ \bibnamefont {Lu}},\ and\ \bibinfo
  {author} {\bibfnamefont {X.~C.}\ \bibnamefont {Xie}},\ }\bibfield  {title}
  {\bibinfo {title} {{Coulomb Drag in Altermagnets}},\ }\href
  {https://doi.org/10.1103/PhysRevLett.134.136301} {\bibfield  {journal}
  {\bibinfo  {journal} {Phys. Rev. Lett.}\ }\textbf {\bibinfo {volume} {134}},\
  \bibinfo {pages} {136301} (\bibinfo {year} {2025})}\BibitemShut {NoStop}%
\bibitem [{\citenamefont {Chen}\ \emph {et~al.}(2025)\citenamefont {Chen},
  \citenamefont {Liu}, \citenamefont {Lu},\ and\ \citenamefont
  {Xie}}]{Chen2025Electrical}%
  \BibitemOpen
  \bibfield  {author} {\bibinfo {author} {\bibfnamefont {Y.}~\bibnamefont
  {Chen}}, \bibinfo {author} {\bibfnamefont {X.}~\bibnamefont {Liu}}, \bibinfo
  {author} {\bibfnamefont {H.-Z.}\ \bibnamefont {Lu}},\ and\ \bibinfo {author}
  {\bibfnamefont {X.~C.}\ \bibnamefont {Xie}},\ }\bibfield  {title} {\bibinfo
  {title} {{Electrical Switching of Altermagnetism}},\ }\href
  {https://doi.org/10.1103/zm5y-vy41} {\bibfield  {journal} {\bibinfo
  {journal} {Phys. Rev. Lett.}\ }\textbf {\bibinfo {volume} {135}},\ \bibinfo
  {pages} {016701} (\bibinfo {year} {2025})}\BibitemShut {NoStop}%
\bibitem [{\citenamefont {Ghorashi}\ \emph {et~al.}(2024)\citenamefont
  {Ghorashi}, \citenamefont {Hughes},\ and\ \citenamefont
  {Cano}}]{Ghorashi2025Altermagnetic}%
  \BibitemOpen
  \bibfield  {author} {\bibinfo {author} {\bibfnamefont {S.~A.~A.}\
  \bibnamefont {Ghorashi}}, \bibinfo {author} {\bibfnamefont {T.~L.}\
  \bibnamefont {Hughes}},\ and\ \bibinfo {author} {\bibfnamefont
  {J.}~\bibnamefont {Cano}},\ }\bibfield  {title} {\bibinfo {title}
  {{Altermagnetic Routes to Majorana Modes in Zero Net Magnetization}},\ }\href
  {https://doi.org/10.1103/PhysRevLett.133.106601} {\bibfield  {journal}
  {\bibinfo  {journal} {Phys. Rev. Lett.}\ }\textbf {\bibinfo {volume} {133}},\
  \bibinfo {pages} {106601} (\bibinfo {year} {2024})}\BibitemShut {NoStop}%
\bibitem [{\citenamefont {Yershov}\ \emph {et~al.}(2025)\citenamefont
  {Yershov}, \citenamefont {Gomonay}, \citenamefont {Sinova}, \citenamefont
  {van~den Brink},\ and\ \citenamefont {Kravchuk}}]{Yershov2025Curvature}%
  \BibitemOpen
  \bibfield  {author} {\bibinfo {author} {\bibfnamefont {K.~V.}\ \bibnamefont
  {Yershov}}, \bibinfo {author} {\bibfnamefont {O.}~\bibnamefont {Gomonay}},
  \bibinfo {author} {\bibfnamefont {J.}~\bibnamefont {Sinova}}, \bibinfo
  {author} {\bibfnamefont {J.}~\bibnamefont {van~den Brink}},\ and\ \bibinfo
  {author} {\bibfnamefont {V.~P.}\ \bibnamefont {Kravchuk}},\ }\bibfield
  {title} {\bibinfo {title} {{Curvature-Induced Magnetization of Altermagnetic
  Films}},\ }\href {https://doi.org/10.1103/PhysRevLett.134.116701} {\bibfield
  {journal} {\bibinfo  {journal} {Phys. Rev. Lett.}\ }\textbf {\bibinfo
  {volume} {134}},\ \bibinfo {pages} {116701} (\bibinfo {year}
  {2025})}\BibitemShut {NoStop}%
\bibitem [{\citenamefont {Ouassou}\ \emph {et~al.}(2023)\citenamefont
  {Ouassou}, \citenamefont {Brataas},\ and\ \citenamefont
  {Linder}}]{Ouassou2023dc}%
  \BibitemOpen
  \bibfield  {author} {\bibinfo {author} {\bibfnamefont {J.~A.}\ \bibnamefont
  {Ouassou}}, \bibinfo {author} {\bibfnamefont {A.}~\bibnamefont {Brataas}},\
  and\ \bibinfo {author} {\bibfnamefont {J.}~\bibnamefont {Linder}},\
  }\bibfield  {title} {\bibinfo {title} {{dc Josephson Effect in
  Altermagnets}},\ }\href {https://doi.org/10.1103/PhysRevLett.131.076003}
  {\bibfield  {journal} {\bibinfo  {journal} {Phys. Rev. Lett.}\ }\textbf
  {\bibinfo {volume} {131}},\ \bibinfo {pages} {076003} (\bibinfo {year}
  {2023})}\BibitemShut {NoStop}%
\bibitem [{\citenamefont {Yuan}\ \emph {et~al.}(2020)\citenamefont {Yuan},
  \citenamefont {Wang}, \citenamefont {Luo}, \citenamefont {Rashba},\ and\
  \citenamefont {Zunger}}]{Yuan2020Giant}%
  \BibitemOpen
  \bibfield  {author} {\bibinfo {author} {\bibfnamefont {L.-D.}\ \bibnamefont
  {Yuan}}, \bibinfo {author} {\bibfnamefont {Z.}~\bibnamefont {Wang}}, \bibinfo
  {author} {\bibfnamefont {J.-W.}\ \bibnamefont {Luo}}, \bibinfo {author}
  {\bibfnamefont {E.~I.}\ \bibnamefont {Rashba}},\ and\ \bibinfo {author}
  {\bibfnamefont {A.}~\bibnamefont {Zunger}},\ }\bibfield  {title} {\bibinfo
  {title} {Giant momentum-dependent spin splitting in centrosymmetric low-$z$
  antiferromagnets},\ }\href {https://doi.org/10.1103/PhysRevB.102.014422}
  {\bibfield  {journal} {\bibinfo  {journal} {Phys. Rev. B}\ }\textbf {\bibinfo
  {volume} {102}},\ \bibinfo {pages} {014422} (\bibinfo {year}
  {2020})}\BibitemShut {NoStop}%
\bibitem [{\citenamefont {Mondal}\ \emph {et~al.}(2025)\citenamefont {Mondal},
  \citenamefont {Pal}, \citenamefont {Saha},\ and\ \citenamefont
  {Nag}}]{Mondal2025Distinguishing}%
  \BibitemOpen
  \bibfield  {author} {\bibinfo {author} {\bibfnamefont {D.}~\bibnamefont
  {Mondal}}, \bibinfo {author} {\bibfnamefont {A.}~\bibnamefont {Pal}},
  \bibinfo {author} {\bibfnamefont {A.}~\bibnamefont {Saha}},\ and\ \bibinfo
  {author} {\bibfnamefont {T.}~\bibnamefont {Nag}},\ }\bibfield  {title}
  {\bibinfo {title} {Distinguishing between topological majorana and trivial
  zero modes via transport and shot noise study in an altermagnet
  heterostructure},\ }\href {https://doi.org/10.1103/PhysRevB.111.L121401}
  {\bibfield  {journal} {\bibinfo  {journal} {Phys. Rev. B}\ }\textbf {\bibinfo
  {volume} {111}},\ \bibinfo {pages} {L121401} (\bibinfo {year}
  {2025})}\BibitemShut {NoStop}%
\bibitem [{\citenamefont {Brekke}\ \emph {et~al.}(2024)\citenamefont {Brekke},
  \citenamefont {Sukhachov}, \citenamefont {Giil}, \citenamefont {Brataas},\
  and\ \citenamefont {Linder}}]{brekke2024minimal}%
  \BibitemOpen
  \bibfield  {author} {\bibinfo {author} {\bibfnamefont {B.}~\bibnamefont
  {Brekke}}, \bibinfo {author} {\bibfnamefont {P.}~\bibnamefont {Sukhachov}},
  \bibinfo {author} {\bibfnamefont {H.~G.}\ \bibnamefont {Giil}}, \bibinfo
  {author} {\bibfnamefont {A.}~\bibnamefont {Brataas}},\ and\ \bibinfo {author}
  {\bibfnamefont {J.}~\bibnamefont {Linder}},\ }\bibfield  {title} {\bibinfo
  {title} {{Minimal models and transport properties of unconventional p-wave
  magnets}},\ }\href {https://doi.org/10.1103/PhysRevLett.133.236703}
  {\bibfield  {journal} {\bibinfo  {journal} {Phys. Rev. Lett.}\ }\textbf
  {\bibinfo {volume} {133}},\ \bibinfo {pages} {236703} (\bibinfo {year}
  {2024})}\BibitemShut {NoStop}%
\bibitem [{\citenamefont {Yu}\ \emph {et~al.}(2025)\citenamefont {Yu},
  \citenamefont {Lyngby}, \citenamefont {Shishidou}, \citenamefont {Roig},
  \citenamefont {Kreisel}, \citenamefont {Weinert}, \citenamefont {Andersen},\
  and\ \citenamefont {Agterberg}}]{yu2025odd}%
  \BibitemOpen
  \bibfield  {author} {\bibinfo {author} {\bibfnamefont {Y.}~\bibnamefont
  {Yu}}, \bibinfo {author} {\bibfnamefont {M.~B.}\ \bibnamefont {Lyngby}},
  \bibinfo {author} {\bibfnamefont {T.}~\bibnamefont {Shishidou}}, \bibinfo
  {author} {\bibfnamefont {M.}~\bibnamefont {Roig}}, \bibinfo {author}
  {\bibfnamefont {A.}~\bibnamefont {Kreisel}}, \bibinfo {author} {\bibfnamefont
  {M.}~\bibnamefont {Weinert}}, \bibinfo {author} {\bibfnamefont {B.~M.}\
  \bibnamefont {Andersen}},\ and\ \bibinfo {author} {\bibfnamefont {D.~F.}\
  \bibnamefont {Agterberg}},\ }\bibfield  {title} {\bibinfo {title}
  {{Odd-parity magnetism driven by antiferromagnetic exchange}},\ }\href
  {https://doi.org/10.1103/zk69-k6b2} {\bibfield  {journal} {\bibinfo
  {journal} {Phys. Rev. Lett.}\ }\textbf {\bibinfo {volume} {135}},\ \bibinfo
  {pages} {046701} (\bibinfo {year} {2025})}\BibitemShut {NoStop}%
\bibitem [{\citenamefont {Song}\ \emph
  {et~al.}(2025{\natexlab{b}})\citenamefont {Song}, \citenamefont
  {Stavri{\'c}}, \citenamefont {Barone}, \citenamefont {Droghetti},
  \citenamefont {Antonenko}, \citenamefont {Venderbos}, \citenamefont
  {Occhialini}, \citenamefont {Ilyas}, \citenamefont {Erge{\c{c}}en},
  \citenamefont {Gedik} \emph {et~al.}}]{song2025electrical}%
  \BibitemOpen
  \bibfield  {author} {\bibinfo {author} {\bibfnamefont {Q.}~\bibnamefont
  {Song}}, \bibinfo {author} {\bibfnamefont {S.}~\bibnamefont {Stavri{\'c}}},
  \bibinfo {author} {\bibfnamefont {P.}~\bibnamefont {Barone}}, \bibinfo
  {author} {\bibfnamefont {A.}~\bibnamefont {Droghetti}}, \bibinfo {author}
  {\bibfnamefont {D.~S.}\ \bibnamefont {Antonenko}}, \bibinfo {author}
  {\bibfnamefont {J.~W.}\ \bibnamefont {Venderbos}}, \bibinfo {author}
  {\bibfnamefont {C.~A.}\ \bibnamefont {Occhialini}}, \bibinfo {author}
  {\bibfnamefont {B.}~\bibnamefont {Ilyas}}, \bibinfo {author} {\bibfnamefont
  {E.}~\bibnamefont {Erge{\c{c}}en}}, \bibinfo {author} {\bibfnamefont
  {N.}~\bibnamefont {Gedik}}, \emph {et~al.},\ }\bibfield  {title} {\bibinfo
  {title} {{Electrical switching of ap-wave magnet}},\ }\href
  {https://doi.org/10.1038/s41586-025-09034-7} {\bibfield  {journal} {\bibinfo
  {journal} {Nature}\ }\textbf {\bibinfo {volume} {642}},\ \bibinfo {pages}
  {64} (\bibinfo {year} {2025}{\natexlab{b}})}\BibitemShut {NoStop}%
\bibitem [{\citenamefont {Hellenes}\ \emph {et~al.}(2023)\citenamefont
  {Hellenes}, \citenamefont {Jungwirth}, \citenamefont {Jaeschke-Ubiergo},
  \citenamefont {Chakraborty}, \citenamefont {Sinova},\ and\ \citenamefont
  {{\v{S}}mejkal}}]{hellenes2023p}%
  \BibitemOpen
  \bibfield  {author} {\bibinfo {author} {\bibfnamefont {A.~B.}\ \bibnamefont
  {Hellenes}}, \bibinfo {author} {\bibfnamefont {T.}~\bibnamefont {Jungwirth}},
  \bibinfo {author} {\bibfnamefont {R.}~\bibnamefont {Jaeschke-Ubiergo}},
  \bibinfo {author} {\bibfnamefont {A.}~\bibnamefont {Chakraborty}}, \bibinfo
  {author} {\bibfnamefont {J.}~\bibnamefont {Sinova}},\ and\ \bibinfo {author}
  {\bibfnamefont {L.}~\bibnamefont {{\v{S}}mejkal}},\ }\bibfield  {title}
  {\bibinfo {title} {{P-wave magnets}},\ }\href@noop {} {\bibfield  {journal}
  {\bibinfo  {journal} {arXiv preprint arXiv:2309.01607}\ } (\bibinfo {year}
  {2023})}\BibitemShut {NoStop}%
\bibitem [{\citenamefont {Hayami}\ \emph {et~al.}(2020)\citenamefont {Hayami},
  \citenamefont {Yanagi},\ and\ \citenamefont
  {Kusunose}}]{Hayami2020Spontaneous}%
  \BibitemOpen
  \bibfield  {author} {\bibinfo {author} {\bibfnamefont {S.}~\bibnamefont
  {Hayami}}, \bibinfo {author} {\bibfnamefont {Y.}~\bibnamefont {Yanagi}},\
  and\ \bibinfo {author} {\bibfnamefont {H.}~\bibnamefont {Kusunose}},\
  }\bibfield  {title} {\bibinfo {title} {Spontaneous antisymmetric spin
  splitting in noncollinear antiferromagnets without spin-orbit coupling},\
  }\href {https://doi.org/10.1103/PhysRevB.101.220403} {\bibfield  {journal}
  {\bibinfo  {journal} {Phys. Rev. B}\ }\textbf {\bibinfo {volume} {101}},\
  \bibinfo {pages} {220403} (\bibinfo {year} {2020})}\BibitemShut {NoStop}%
\bibitem [{\citenamefont {Yamada}\ \emph {et~al.}(2025)\citenamefont {Yamada},
  \citenamefont {Birch}, \citenamefont {Baral}, \citenamefont {Okumura},
  \citenamefont {Nakano}, \citenamefont {Gao}, \citenamefont {Ezawa},
  \citenamefont {Nomoto}, \citenamefont {Masell}, \citenamefont {Ishihara}
  \emph {et~al.}}]{yamada2025metallic}%
  \BibitemOpen
  \bibfield  {author} {\bibinfo {author} {\bibfnamefont {R.}~\bibnamefont
  {Yamada}}, \bibinfo {author} {\bibfnamefont {M.~T.}\ \bibnamefont {Birch}},
  \bibinfo {author} {\bibfnamefont {P.~R.}\ \bibnamefont {Baral}}, \bibinfo
  {author} {\bibfnamefont {S.}~\bibnamefont {Okumura}}, \bibinfo {author}
  {\bibfnamefont {R.}~\bibnamefont {Nakano}}, \bibinfo {author} {\bibfnamefont
  {S.}~\bibnamefont {Gao}}, \bibinfo {author} {\bibfnamefont {M.}~\bibnamefont
  {Ezawa}}, \bibinfo {author} {\bibfnamefont {T.}~\bibnamefont {Nomoto}},
  \bibinfo {author} {\bibfnamefont {J.}~\bibnamefont {Masell}}, \bibinfo
  {author} {\bibfnamefont {Y.}~\bibnamefont {Ishihara}}, \emph {et~al.},\
  }\bibfield  {title} {\bibinfo {title} {A metallic p-wave magnet with
  commensurate spin helix},\ }\href
  {https://doi.org/https://doi.org/10.1038/s41586-025-09633-4} {\bibfield
  {journal} {\bibinfo  {journal} {Nature}\ }\textbf {\bibinfo {volume} {646}},\
  \bibinfo {pages} {837} (\bibinfo {year} {2025})}\BibitemShut {NoStop}%
\bibitem [{\citenamefont {Wang}\ \emph {et~al.}(2013)\citenamefont {Wang},
  \citenamefont {Steinberg}, \citenamefont {Jarillo-Herrero},\ and\
  \citenamefont {Gedik}}]{wang2013observation}%
  \BibitemOpen
  \bibfield  {author} {\bibinfo {author} {\bibfnamefont {Y.}~\bibnamefont
  {Wang}}, \bibinfo {author} {\bibfnamefont {H.}~\bibnamefont {Steinberg}},
  \bibinfo {author} {\bibfnamefont {P.}~\bibnamefont {Jarillo-Herrero}},\ and\
  \bibinfo {author} {\bibfnamefont {N.}~\bibnamefont {Gedik}},\ }\bibfield
  {title} {\bibinfo {title} {{Observation of Floquet-Bloch states on the
  surface of a topological insulator}},\ }\href
  {https://doi.org/10.1126/science.1239834} {\bibfield  {journal} {\bibinfo
  {journal} {Science}\ }\textbf {\bibinfo {volume} {342}},\ \bibinfo {pages}
  {453} (\bibinfo {year} {2013})}\BibitemShut {NoStop}%
\bibitem [{\citenamefont {McIver}\ \emph {et~al.}(2020)\citenamefont {McIver},
  \citenamefont {Schulte}, \citenamefont {Stein}, \citenamefont {Matsuyama},
  \citenamefont {Jotzu}, \citenamefont {Meier},\ and\ \citenamefont
  {Cavalleri}}]{mciver2020light}%
  \BibitemOpen
  \bibfield  {author} {\bibinfo {author} {\bibfnamefont {J.~W.}\ \bibnamefont
  {McIver}}, \bibinfo {author} {\bibfnamefont {B.}~\bibnamefont {Schulte}},
  \bibinfo {author} {\bibfnamefont {F.-U.}\ \bibnamefont {Stein}}, \bibinfo
  {author} {\bibfnamefont {T.}~\bibnamefont {Matsuyama}}, \bibinfo {author}
  {\bibfnamefont {G.}~\bibnamefont {Jotzu}}, \bibinfo {author} {\bibfnamefont
  {G.}~\bibnamefont {Meier}},\ and\ \bibinfo {author} {\bibfnamefont
  {A.}~\bibnamefont {Cavalleri}},\ }\bibfield  {title} {\bibinfo {title}
  {{Light-induced anomalous Hall effect in graphene}},\ }\href
  {https://doi.org/10.1038/s41567-019-0698-y} {\bibfield  {journal} {\bibinfo
  {journal} {Nat. Phys.}\ }\textbf {\bibinfo {volume} {16}},\ \bibinfo {pages}
  {38} (\bibinfo {year} {2020})}\BibitemShut {NoStop}%
\bibitem [{\citenamefont {Zhu}\ \emph {et~al.}(2023)\citenamefont {Zhu},
  \citenamefont {Wang},\ and\ \citenamefont {Zhang}}]{zhu2023floquet}%
  \BibitemOpen
  \bibfield  {author} {\bibinfo {author} {\bibfnamefont {T.}~\bibnamefont
  {Zhu}}, \bibinfo {author} {\bibfnamefont {H.}~\bibnamefont {Wang}},\ and\
  \bibinfo {author} {\bibfnamefont {H.}~\bibnamefont {Zhang}},\ }\bibfield
  {title} {\bibinfo {title} {{Floquet engineering of magnetic topological
  insulator MnBi$_2$Te$_4$ films}},\ }\href
  {https://doi.org/10.1103/PhysRevB.107.085151} {\bibfield  {journal} {\bibinfo
   {journal} {Phys. Rev. B}\ }\textbf {\bibinfo {volume} {107}},\ \bibinfo
  {pages} {085151} (\bibinfo {year} {2023})}\BibitemShut {NoStop}%
\bibitem [{\citenamefont {Zhou}\ \emph {et~al.}(2023)\citenamefont {Zhou},
  \citenamefont {Bao}, \citenamefont {Fan}, \citenamefont {Zhou}, \citenamefont
  {Gao}, \citenamefont {Zhong}, \citenamefont {Lin}, \citenamefont {Liu},
  \citenamefont {Yu}, \citenamefont {Tang} \emph
  {et~al.}}]{zhou2023pseudospin}%
  \BibitemOpen
  \bibfield  {author} {\bibinfo {author} {\bibfnamefont {S.}~\bibnamefont
  {Zhou}}, \bibinfo {author} {\bibfnamefont {C.}~\bibnamefont {Bao}}, \bibinfo
  {author} {\bibfnamefont {B.}~\bibnamefont {Fan}}, \bibinfo {author}
  {\bibfnamefont {H.}~\bibnamefont {Zhou}}, \bibinfo {author} {\bibfnamefont
  {Q.}~\bibnamefont {Gao}}, \bibinfo {author} {\bibfnamefont {H.}~\bibnamefont
  {Zhong}}, \bibinfo {author} {\bibfnamefont {T.}~\bibnamefont {Lin}}, \bibinfo
  {author} {\bibfnamefont {H.}~\bibnamefont {Liu}}, \bibinfo {author}
  {\bibfnamefont {P.}~\bibnamefont {Yu}}, \bibinfo {author} {\bibfnamefont
  {P.}~\bibnamefont {Tang}}, \emph {et~al.},\ }\bibfield  {title} {\bibinfo
  {title} {{Pseudospin-selective Floquet band engineering in black
  phosphorus}},\ }\href {https://doi.org/10.1038/s41586-022-05610-3} {\bibfield
   {journal} {\bibinfo  {journal} {Nature}\ }\textbf {\bibinfo {volume}
  {614}},\ \bibinfo {pages} {75} (\bibinfo {year} {2023})}\BibitemShut
  {NoStop}%
\bibitem [{\citenamefont {Ning}\ \emph {et~al.}(2024)\citenamefont {Ning},
  \citenamefont {Ma}, \citenamefont {Zeng}, \citenamefont {Xu},\ and\
  \citenamefont {Wang}}]{ning2024flexible}%
  \BibitemOpen
  \bibfield  {author} {\bibinfo {author} {\bibfnamefont {Z.}~\bibnamefont
  {Ning}}, \bibinfo {author} {\bibfnamefont {D.-S.}\ \bibnamefont {Ma}},
  \bibinfo {author} {\bibfnamefont {J.}~\bibnamefont {Zeng}}, \bibinfo {author}
  {\bibfnamefont {D.-H.}\ \bibnamefont {Xu}},\ and\ \bibinfo {author}
  {\bibfnamefont {R.}~\bibnamefont {Wang}},\ }\bibfield  {title} {\bibinfo
  {title} {{Flexible Control of Chiral Superconductivity in Optically Driven
  Nodal Point Superconductors with Antiferromagnetism}},\ }\href
  {https://doi.org/10.1103/PhysRevLett.133.246606} {\bibfield  {journal}
  {\bibinfo  {journal} {Phys. Rev. Lett.}\ }\textbf {\bibinfo {volume} {133}},\
  \bibinfo {pages} {246606} (\bibinfo {year} {2024})}\BibitemShut {NoStop}%
\bibitem [{\citenamefont {Liu}\ \emph {et~al.}(2018)\citenamefont {Liu},
  \citenamefont {Sun}, \citenamefont {Cheng}, \citenamefont {Liu},\ and\
  \citenamefont {Meng}}]{liu2018photoinduced}%
  \BibitemOpen
  \bibfield  {author} {\bibinfo {author} {\bibfnamefont {H.}~\bibnamefont
  {Liu}}, \bibinfo {author} {\bibfnamefont {J.-T.}\ \bibnamefont {Sun}},
  \bibinfo {author} {\bibfnamefont {C.}~\bibnamefont {Cheng}}, \bibinfo
  {author} {\bibfnamefont {F.}~\bibnamefont {Liu}},\ and\ \bibinfo {author}
  {\bibfnamefont {S.}~\bibnamefont {Meng}},\ }\bibfield  {title} {\bibinfo
  {title} {{Photoinduced nonequilibrium topological states in strained black
  phosphorus}},\ }\href {https://doi.org/10.1103/PhysRevLett.120.237403}
  {\bibfield  {journal} {\bibinfo  {journal} {Phys. Rev. Lett.}\ }\textbf
  {\bibinfo {volume} {120}},\ \bibinfo {pages} {237403} (\bibinfo {year}
  {2018})}\BibitemShut {NoStop}%
\bibitem [{\citenamefont {Zhang}\ and\ \citenamefont
  {Das~Sarma}(2021)}]{Zhang2021Anomalous}%
  \BibitemOpen
  \bibfield  {author} {\bibinfo {author} {\bibfnamefont {R.-X.}\ \bibnamefont
  {Zhang}}\ and\ \bibinfo {author} {\bibfnamefont {S.}~\bibnamefont
  {Das~Sarma}},\ }\bibfield  {title} {\bibinfo {title} {{Anomalous Floquet
  Chiral Topological Superconductivity in a Topological Insulator Sandwich
  Structure}},\ }\href {https://doi.org/10.1103/PhysRevLett.127.067001}
  {\bibfield  {journal} {\bibinfo  {journal} {Phys. Rev. Lett.}\ }\textbf
  {\bibinfo {volume} {127}},\ \bibinfo {pages} {067001} (\bibinfo {year}
  {2021})}\BibitemShut {NoStop}%
\bibitem [{\citenamefont {Grushin}\ \emph {et~al.}(2014)\citenamefont
  {Grushin}, \citenamefont {G\'omez-Le\'on},\ and\ \citenamefont
  {Neupert}}]{Grushin2014Floquet}%
  \BibitemOpen
  \bibfield  {author} {\bibinfo {author} {\bibfnamefont {A.~G.}\ \bibnamefont
  {Grushin}}, \bibinfo {author} {\bibfnamefont {A.}~\bibnamefont
  {G\'omez-Le\'on}},\ and\ \bibinfo {author} {\bibfnamefont {T.}~\bibnamefont
  {Neupert}},\ }\bibfield  {title} {\bibinfo {title} {{Floquet Fractional Chern
  Insulators}},\ }\href {https://doi.org/10.1103/PhysRevLett.112.156801}
  {\bibfield  {journal} {\bibinfo  {journal} {Phys. Rev. Lett.}\ }\textbf
  {\bibinfo {volume} {112}},\ \bibinfo {pages} {156801} (\bibinfo {year}
  {2014})}\BibitemShut {NoStop}%
\bibitem [{\citenamefont {Yan}\ and\ \citenamefont
  {Wang}(2016)}]{Yan2016Tunable}%
  \BibitemOpen
  \bibfield  {author} {\bibinfo {author} {\bibfnamefont {Z.}~\bibnamefont
  {Yan}}\ and\ \bibinfo {author} {\bibfnamefont {Z.}~\bibnamefont {Wang}},\
  }\bibfield  {title} {\bibinfo {title} {{Tunable Weyl Points in Periodically
  Driven Nodal Line Semimetals}},\ }\href
  {https://doi.org/10.1103/PhysRevLett.117.087402} {\bibfield  {journal}
  {\bibinfo  {journal} {Phys. Rev. Lett.}\ }\textbf {\bibinfo {volume} {117}},\
  \bibinfo {pages} {087402} (\bibinfo {year} {2016})}\BibitemShut {NoStop}%
\bibitem [{\citenamefont {H{\"u}bener}\ \emph {et~al.}(2017)\citenamefont
  {H{\"u}bener}, \citenamefont {Sentef}, \citenamefont {De~Giovannini},
  \citenamefont {Kemper},\ and\ \citenamefont {Rubio}}]{hubener2017creating}%
  \BibitemOpen
  \bibfield  {author} {\bibinfo {author} {\bibfnamefont {H.}~\bibnamefont
  {H{\"u}bener}}, \bibinfo {author} {\bibfnamefont {M.~A.}\ \bibnamefont
  {Sentef}}, \bibinfo {author} {\bibfnamefont {U.}~\bibnamefont
  {De~Giovannini}}, \bibinfo {author} {\bibfnamefont {A.~F.}\ \bibnamefont
  {Kemper}},\ and\ \bibinfo {author} {\bibfnamefont {A.}~\bibnamefont
  {Rubio}},\ }\bibfield  {title} {\bibinfo {title} {{Creating stable
  Floquet--Weyl semimetals by laser-driving of 3D Dirac materials}},\ }\href
  {https://doi.org/10.1038/ncomms13940} {\bibfield  {journal} {\bibinfo
  {journal} {Nat. Commun.}\ }\textbf {\bibinfo {volume} {8}},\ \bibinfo {pages}
  {13940} (\bibinfo {year} {2017})}\BibitemShut {NoStop}%
\bibitem [{\citenamefont {Bao}\ \emph {et~al.}(2024)\citenamefont {Bao},
  \citenamefont {Sch{\"u}ler}, \citenamefont {Xiao}, \citenamefont {Wang},
  \citenamefont {Zhong}, \citenamefont {Lin}, \citenamefont {Cai},
  \citenamefont {Sheng}, \citenamefont {Tang}, \citenamefont {Zhang} \emph
  {et~al.}}]{bao2024manipulating}%
  \BibitemOpen
  \bibfield  {author} {\bibinfo {author} {\bibfnamefont {C.}~\bibnamefont
  {Bao}}, \bibinfo {author} {\bibfnamefont {M.}~\bibnamefont {Sch{\"u}ler}},
  \bibinfo {author} {\bibfnamefont {T.}~\bibnamefont {Xiao}}, \bibinfo {author}
  {\bibfnamefont {F.}~\bibnamefont {Wang}}, \bibinfo {author} {\bibfnamefont
  {H.}~\bibnamefont {Zhong}}, \bibinfo {author} {\bibfnamefont
  {T.}~\bibnamefont {Lin}}, \bibinfo {author} {\bibfnamefont {X.}~\bibnamefont
  {Cai}}, \bibinfo {author} {\bibfnamefont {T.}~\bibnamefont {Sheng}}, \bibinfo
  {author} {\bibfnamefont {X.}~\bibnamefont {Tang}}, \bibinfo {author}
  {\bibfnamefont {H.}~\bibnamefont {Zhang}}, \emph {et~al.},\ }\bibfield
  {title} {\bibinfo {title} {{Manipulating the symmetry of photon-dressed
  electronic states}},\ }\href {https://doi.org/10.1038/s41467-024-54760-7}
  {\bibfield  {journal} {\bibinfo  {journal} {Nat. Commun.}\ }\textbf {\bibinfo
  {volume} {15}},\ \bibinfo {pages} {10535} (\bibinfo {year}
  {2024})}\BibitemShut {NoStop}%
\bibitem [{\citenamefont {Liu}\ \emph {et~al.}(2025)\citenamefont {Liu},
  \citenamefont {Yang}, \citenamefont {Gaertner}, \citenamefont {Huckabee},
  \citenamefont {Suslov}, \citenamefont {Refael}, \citenamefont {Nathan},
  \citenamefont {Lewandowski}, \citenamefont {Foa~Torres}, \citenamefont {Esin}
  \emph {et~al.}}]{liu2025signatures}%
  \BibitemOpen
  \bibfield  {author} {\bibinfo {author} {\bibfnamefont {Y.}~\bibnamefont
  {Liu}}, \bibinfo {author} {\bibfnamefont {C.}~\bibnamefont {Yang}}, \bibinfo
  {author} {\bibfnamefont {G.}~\bibnamefont {Gaertner}}, \bibinfo {author}
  {\bibfnamefont {J.}~\bibnamefont {Huckabee}}, \bibinfo {author}
  {\bibfnamefont {A.~V.}\ \bibnamefont {Suslov}}, \bibinfo {author}
  {\bibfnamefont {G.}~\bibnamefont {Refael}}, \bibinfo {author} {\bibfnamefont
  {F.}~\bibnamefont {Nathan}}, \bibinfo {author} {\bibfnamefont
  {C.}~\bibnamefont {Lewandowski}}, \bibinfo {author} {\bibfnamefont {L.~E.}\
  \bibnamefont {Foa~Torres}}, \bibinfo {author} {\bibfnamefont
  {I.}~\bibnamefont {Esin}}, \emph {et~al.},\ }\bibfield  {title} {\bibinfo
  {title} {{Signatures of Floquet electronic steady states in graphene under
  continuous-wave mid-infrared irradiation}},\ }\href
  {https://doi.org/10.1038/s41467-025-57335-2} {\bibfield  {journal} {\bibinfo
  {journal} {Nat. Commun.}\ }\textbf {\bibinfo {volume} {16}},\ \bibinfo
  {pages} {2057} (\bibinfo {year} {2025})}\BibitemShut {NoStop}%
\bibitem [{\citenamefont {Fan}\ \emph {et~al.}(2025)\citenamefont {Fan},
  \citenamefont {De~Giovannini}, \citenamefont {H{\"u}bener}, \citenamefont
  {Zhou}, \citenamefont {Duan}, \citenamefont {Rubio},\ and\ \citenamefont
  {Tang}}]{fan2025floquet}%
  \BibitemOpen
  \bibfield  {author} {\bibinfo {author} {\bibfnamefont {B.}~\bibnamefont
  {Fan}}, \bibinfo {author} {\bibfnamefont {U.}~\bibnamefont {De~Giovannini}},
  \bibinfo {author} {\bibfnamefont {H.}~\bibnamefont {H{\"u}bener}}, \bibinfo
  {author} {\bibfnamefont {S.}~\bibnamefont {Zhou}}, \bibinfo {author}
  {\bibfnamefont {W.}~\bibnamefont {Duan}}, \bibinfo {author} {\bibfnamefont
  {A.}~\bibnamefont {Rubio}},\ and\ \bibinfo {author} {\bibfnamefont
  {P.}~\bibnamefont {Tang}},\ }\bibfield  {title} {\bibinfo {title} {{Floquet
  optical selection rules in black phosphorus}},\ }\href@noop {} {\bibfield
  {journal} {\bibinfo  {journal} {arXiv preprint arXiv:2501.15703}\ } (\bibinfo
  {year} {2025})}\BibitemShut {NoStop}%
\bibitem [{\citenamefont {Ghorashi}\ and\ \citenamefont
  {Li}(2025)}]{ghorashi2025dynamical}%
  \BibitemOpen
  \bibfield  {author} {\bibinfo {author} {\bibfnamefont {S.~A.~A.}\
  \bibnamefont {Ghorashi}}\ and\ \bibinfo {author} {\bibfnamefont
  {Q.}~\bibnamefont {Li}},\ }\bibfield  {title} {\bibinfo {title} {Dynamical
  generation of higher-order spin-orbit couplings, topology and persistent spin
  texture in light-irradiated altermagnets},\ }\href@noop {} {\bibfield
  {journal} {\bibinfo  {journal} {arXiv preprint arXiv:2504.00122}\ } (\bibinfo
  {year} {2025})}\BibitemShut {NoStop}%
\bibitem [{\citenamefont {Jiang}\ \emph {et~al.}(2024)\citenamefont {Jiang},
  \citenamefont {Song}, \citenamefont {Zhu}, \citenamefont {Fang},
  \citenamefont {Weng}, \citenamefont {Liu}, \citenamefont {Yang},\ and\
  \citenamefont {Fang}}]{jiang2024enumeration}%
  \BibitemOpen
  \bibfield  {author} {\bibinfo {author} {\bibfnamefont {Y.}~\bibnamefont
  {Jiang}}, \bibinfo {author} {\bibfnamefont {Z.}~\bibnamefont {Song}},
  \bibinfo {author} {\bibfnamefont {T.}~\bibnamefont {Zhu}}, \bibinfo {author}
  {\bibfnamefont {Z.}~\bibnamefont {Fang}}, \bibinfo {author} {\bibfnamefont
  {H.}~\bibnamefont {Weng}}, \bibinfo {author} {\bibfnamefont {Z.-X.}\
  \bibnamefont {Liu}}, \bibinfo {author} {\bibfnamefont {J.}~\bibnamefont
  {Yang}},\ and\ \bibinfo {author} {\bibfnamefont {C.}~\bibnamefont {Fang}},\
  }\bibfield  {title} {\bibinfo {title} {{Enumeration of spin-space groups:
  Toward a complete description of symmetries of magnetic orders}},\ }\href
  {https://doi.org/10.1103/PhysRevX.14.031039} {\bibfield  {journal} {\bibinfo
  {journal} {Phys. Rev. X}\ }\textbf {\bibinfo {volume} {14}},\ \bibinfo
  {pages} {031039} (\bibinfo {year} {2024})}\BibitemShut {NoStop}%
\bibitem [{\citenamefont {Chen}\ \emph {et~al.}(2024)\citenamefont {Chen},
  \citenamefont {Ren}, \citenamefont {Zhu}, \citenamefont {Yu}, \citenamefont
  {Zhang}, \citenamefont {Liu}, \citenamefont {Li}, \citenamefont {Liu},
  \citenamefont {Li},\ and\ \citenamefont {Liu}}]{chen2024enumeration}%
  \BibitemOpen
  \bibfield  {author} {\bibinfo {author} {\bibfnamefont {X.}~\bibnamefont
  {Chen}}, \bibinfo {author} {\bibfnamefont {J.}~\bibnamefont {Ren}}, \bibinfo
  {author} {\bibfnamefont {Y.}~\bibnamefont {Zhu}}, \bibinfo {author}
  {\bibfnamefont {Y.}~\bibnamefont {Yu}}, \bibinfo {author} {\bibfnamefont
  {A.}~\bibnamefont {Zhang}}, \bibinfo {author} {\bibfnamefont
  {P.}~\bibnamefont {Liu}}, \bibinfo {author} {\bibfnamefont {J.}~\bibnamefont
  {Li}}, \bibinfo {author} {\bibfnamefont {Y.}~\bibnamefont {Liu}}, \bibinfo
  {author} {\bibfnamefont {C.}~\bibnamefont {Li}},\ and\ \bibinfo {author}
  {\bibfnamefont {Q.}~\bibnamefont {Liu}},\ }\bibfield  {title} {\bibinfo
  {title} {{Enumeration and representation theory of spin space groups}},\
  }\href {https://doi.org/10.1103/PhysRevX.14.031038} {\bibfield  {journal}
  {\bibinfo  {journal} {Phys. Rev. X}\ }\textbf {\bibinfo {volume} {14}},\
  \bibinfo {pages} {031038} (\bibinfo {year} {2024})}\BibitemShut {NoStop}%
\bibitem [{\citenamefont {Xiao}\ \emph {et~al.}(2024)\citenamefont {Xiao},
  \citenamefont {Zhao}, \citenamefont {Li}, \citenamefont {Shindou},\ and\
  \citenamefont {Song}}]{xiao2024spin}%
  \BibitemOpen
  \bibfield  {author} {\bibinfo {author} {\bibfnamefont {Z.}~\bibnamefont
  {Xiao}}, \bibinfo {author} {\bibfnamefont {J.}~\bibnamefont {Zhao}}, \bibinfo
  {author} {\bibfnamefont {Y.}~\bibnamefont {Li}}, \bibinfo {author}
  {\bibfnamefont {R.}~\bibnamefont {Shindou}},\ and\ \bibinfo {author}
  {\bibfnamefont {Z.-D.}\ \bibnamefont {Song}},\ }\bibfield  {title} {\bibinfo
  {title} {{Spin space groups: Full classification and applications}},\ }\href
  {https://doi.org/10.1103/PhysRevX.14.031037} {\bibfield  {journal} {\bibinfo
  {journal} {Phys. Rev. X}\ }\textbf {\bibinfo {volume} {14}},\ \bibinfo
  {pages} {031037} (\bibinfo {year} {2024})}\BibitemShut {NoStop}%
\bibitem [{\citenamefont {Kitagawa}\ \emph {et~al.}(2011)\citenamefont
  {Kitagawa}, \citenamefont {Oka}, \citenamefont {Brataas}, \citenamefont
  {Fu},\ and\ \citenamefont {Demler}}]{Kitagawa2011Transport}%
  \BibitemOpen
  \bibfield  {author} {\bibinfo {author} {\bibfnamefont {T.}~\bibnamefont
  {Kitagawa}}, \bibinfo {author} {\bibfnamefont {T.}~\bibnamefont {Oka}},
  \bibinfo {author} {\bibfnamefont {A.}~\bibnamefont {Brataas}}, \bibinfo
  {author} {\bibfnamefont {L.}~\bibnamefont {Fu}},\ and\ \bibinfo {author}
  {\bibfnamefont {E.}~\bibnamefont {Demler}},\ }\bibfield  {title} {\bibinfo
  {title} {{Transport properties of nonequilibrium systems under the
  application of light: Photoinduced quantum Hall insulators without Landau
  levels}},\ }\href {https://doi.org/10.1103/PhysRevB.84.235108} {\bibfield
  {journal} {\bibinfo  {journal} {Phys. Rev. B}\ }\textbf {\bibinfo {volume}
  {84}},\ \bibinfo {pages} {235108} (\bibinfo {year} {2011})}\BibitemShut
  {NoStop}%
\bibitem [{\citenamefont {Mikami}\ \emph {et~al.}(2016)\citenamefont {Mikami},
  \citenamefont {Kitamura}, \citenamefont {Yasuda}, \citenamefont {Tsuji},
  \citenamefont {Oka},\ and\ \citenamefont {Aoki}}]{Mikami2016Brillouin}%
  \BibitemOpen
  \bibfield  {author} {\bibinfo {author} {\bibfnamefont {T.}~\bibnamefont
  {Mikami}}, \bibinfo {author} {\bibfnamefont {S.}~\bibnamefont {Kitamura}},
  \bibinfo {author} {\bibfnamefont {K.}~\bibnamefont {Yasuda}}, \bibinfo
  {author} {\bibfnamefont {N.}~\bibnamefont {Tsuji}}, \bibinfo {author}
  {\bibfnamefont {T.}~\bibnamefont {Oka}},\ and\ \bibinfo {author}
  {\bibfnamefont {H.}~\bibnamefont {Aoki}},\ }\bibfield  {title} {\bibinfo
  {title} {{Brillouin-Wigner theory for high-frequency expansion in
  periodically driven systems: Application to Floquet topological
  insulators}},\ }\href {https://doi.org/10.1103/PhysRevB.93.144307} {\bibfield
   {journal} {\bibinfo  {journal} {Phys. Rev. B}\ }\textbf {\bibinfo {volume}
  {93}},\ \bibinfo {pages} {144307} (\bibinfo {year} {2016})}\BibitemShut
  {NoStop}%
\bibitem [{\citenamefont {Bukov}\ \emph {et~al.}(2015)\citenamefont {Bukov},
  \citenamefont {D'Alessio},\ and\ \citenamefont
  {Polkovnikov}}]{Marin2015Universal}%
  \BibitemOpen
  \bibfield  {author} {\bibinfo {author} {\bibfnamefont {M.}~\bibnamefont
  {Bukov}}, \bibinfo {author} {\bibfnamefont {L.}~\bibnamefont {D'Alessio}},\
  and\ \bibinfo {author} {\bibfnamefont {A.}~\bibnamefont {Polkovnikov}},\
  }\bibfield  {title} {\bibinfo {title} {{Universal high-frequency behavior of
  periodically driven systems: from dynamical stabilization to Floquet
  engineering}},\ }\href {https://doi.org/10.1080/00018732.2015.1055918}
  {\bibfield  {journal} {\bibinfo  {journal} {Adv. Phys.}\ }\textbf {\bibinfo
  {volume} {64}},\ \bibinfo {pages} {139} (\bibinfo {year} {2015})}\BibitemShut
  {NoStop}%
\bibitem [{\citenamefont {Eckardt}\ and\ \citenamefont
  {Anisimovas}(2015)}]{Eckardt2015High}%
  \BibitemOpen
  \bibfield  {author} {\bibinfo {author} {\bibfnamefont {A.}~\bibnamefont
  {Eckardt}}\ and\ \bibinfo {author} {\bibfnamefont {E.}~\bibnamefont
  {Anisimovas}},\ }\bibfield  {title} {\bibinfo {title} {{High-frequency
  approximation for periodically driven quantum systems from a Floquet-space
  perspective}},\ }\href {https://doi.org/10.1088/1367-2630/17/9/093039}
  {\bibfield  {journal} {\bibinfo  {journal} {New J. Phys.}\ }\textbf {\bibinfo
  {volume} {17}},\ \bibinfo {pages} {093039} (\bibinfo {year}
  {2015})}\BibitemShut {NoStop}%
\bibitem [{SM()}]{SM}%
  \BibitemOpen
  \href@noop {} {}\bibinfo {note} {See the Supplementary Material, which
  includes Refs. \cite{Kresse1996vasp, Kresse1999vasp,Perdew1996PBE,
  Blochl1994PBE,marzari1997maximally,
  souza2001maximally,Pizzi2020wannier90,WU2018WannierTools,zhi2022wannsymm,merboldt2025observation,sentef2015theory,freericks2009theoretical,farrell2016dirac},
  for the method and first-principles calculations for Floquet f-wave Chern
  insulator phase in MnPSe$_{3}$ and p-wave magnet in strained
  MnPSe$_{3}$.}\BibitemShut {Stop}%
\bibitem [{\citenamefont {Chittari}\ \emph {et~al.}(2016)\citenamefont
  {Chittari}, \citenamefont {Park}, \citenamefont {Lee}, \citenamefont {Han},
  \citenamefont {MacDonald}, \citenamefont {Hwang},\ and\ \citenamefont
  {Jung}}]{Chittari2016Electronic}%
  \BibitemOpen
  \bibfield  {author} {\bibinfo {author} {\bibfnamefont {B.~L.}\ \bibnamefont
  {Chittari}}, \bibinfo {author} {\bibfnamefont {Y.}~\bibnamefont {Park}},
  \bibinfo {author} {\bibfnamefont {D.}~\bibnamefont {Lee}}, \bibinfo {author}
  {\bibfnamefont {M.}~\bibnamefont {Han}}, \bibinfo {author} {\bibfnamefont
  {A.~H.}\ \bibnamefont {MacDonald}}, \bibinfo {author} {\bibfnamefont
  {E.}~\bibnamefont {Hwang}},\ and\ \bibinfo {author} {\bibfnamefont
  {J.}~\bibnamefont {Jung}},\ }\bibfield  {title} {\bibinfo {title}
  {{Electronic and magnetic properties of single-layer $M\mathrm{P}{X}_{3}$
  metal phosphorous trichalcogenides}},\ }\href
  {https://doi.org/10.1103/PhysRevB.94.184428} {\bibfield  {journal} {\bibinfo
  {journal} {Phys. Rev. B}\ }\textbf {\bibinfo {volume} {94}},\ \bibinfo
  {pages} {184428} (\bibinfo {year} {2016})}\BibitemShut {NoStop}%
\bibitem [{\citenamefont {Huang}\ \emph {et~al.}(2025)\citenamefont {Huang},
  \citenamefont {Qin}, \citenamefont {Zhan}, \citenamefont {Xu}, \citenamefont
  {Wang} \emph {et~al.}}]{huang2025light}%
  \BibitemOpen
  \bibfield  {author} {\bibinfo {author} {\bibfnamefont {S.}~\bibnamefont
  {Huang}}, \bibinfo {author} {\bibfnamefont {Z.}~\bibnamefont {Qin}}, \bibinfo
  {author} {\bibfnamefont {F.}~\bibnamefont {Zhan}}, \bibinfo {author}
  {\bibfnamefont {D.-H.}\ \bibnamefont {Xu}}, \bibinfo {author} {\bibfnamefont
  {R.}~\bibnamefont {Wang}}, \emph {et~al.},\ }\bibfield  {title} {\bibinfo
  {title} {{Light-induced Odd-parity Magnetism in Conventional Collinear
  Antiferromagnets}},\ }\href@noop {} {\bibfield  {journal} {\bibinfo
  {journal} {arXiv preprint arXiv:2507.20705}\ } (\bibinfo {year}
  {2025})}\BibitemShut {NoStop}%
\bibitem [{\citenamefont {Li}\ \emph {et~al.}(2025)\citenamefont {Li},
  \citenamefont {Shao},\ and\ \citenamefont {Kovalev}}]{li2025floquet}%
  \BibitemOpen
  \bibfield  {author} {\bibinfo {author} {\bibfnamefont {B.}~\bibnamefont
  {Li}}, \bibinfo {author} {\bibfnamefont {D.-F.}\ \bibnamefont {Shao}},\ and\
  \bibinfo {author} {\bibfnamefont {A.~A.}\ \bibnamefont {Kovalev}},\
  }\bibfield  {title} {\bibinfo {title} {{Floquet Spin Splitting and Spin
  Generation in Antiferromagnets}},\ }\href@noop {} {\bibfield  {journal}
  {\bibinfo  {journal} {arXiv preprint arXiv:2507.22884}\ } (\bibinfo {year}
  {2025})}\BibitemShut {NoStop}%
\bibitem [{\citenamefont {Kresse}\ and\ \citenamefont
  {Furthm\"uller}(1996)}]{Kresse1996vasp}%
  \BibitemOpen
  \bibfield  {author} {\bibinfo {author} {\bibfnamefont {G.}~\bibnamefont
  {Kresse}}\ and\ \bibinfo {author} {\bibfnamefont {J.}~\bibnamefont
  {Furthm\"uller}},\ }\bibfield  {title} {\bibinfo {title} {{Efficient
  iterative schemes for ab initio total-energy calculations using a plane-wave
  basis set}},\ }\href {https://doi.org/10.1103/PhysRevB.54.11169} {\bibfield
  {journal} {\bibinfo  {journal} {Phys. Rev. B}\ }\textbf {\bibinfo {volume}
  {54}},\ \bibinfo {pages} {11169} (\bibinfo {year} {1996})}\BibitemShut
  {NoStop}%
\bibitem [{\citenamefont {Kresse}\ and\ \citenamefont
  {Joubert}(1999)}]{Kresse1999vasp}%
  \BibitemOpen
  \bibfield  {author} {\bibinfo {author} {\bibfnamefont {G.}~\bibnamefont
  {Kresse}}\ and\ \bibinfo {author} {\bibfnamefont {D.}~\bibnamefont
  {Joubert}},\ }\bibfield  {title} {\bibinfo {title} {{From ultrasoft
  pseudopotentials to the projector augmented-wave method}},\ }\href
  {https://doi.org/10.1103/PhysRevB.59.1758} {\bibfield  {journal} {\bibinfo
  {journal} {Phys. Rev. B}\ }\textbf {\bibinfo {volume} {59}},\ \bibinfo
  {pages} {1758} (\bibinfo {year} {1999})}\BibitemShut {NoStop}%
\bibitem [{\citenamefont {Perdew}\ \emph {et~al.}(1996)\citenamefont {Perdew},
  \citenamefont {Burke},\ and\ \citenamefont {Ernzerhof}}]{Perdew1996PBE}%
  \BibitemOpen
  \bibfield  {author} {\bibinfo {author} {\bibfnamefont {J.~P.}\ \bibnamefont
  {Perdew}}, \bibinfo {author} {\bibfnamefont {K.}~\bibnamefont {Burke}},\ and\
  \bibinfo {author} {\bibfnamefont {M.}~\bibnamefont {Ernzerhof}},\ }\bibfield
  {title} {\bibinfo {title} {{Generalized Gradient Approximation Made
  Simple}},\ }\href {https://doi.org/10.1103/PhysRevLett.77.3865} {\bibfield
  {journal} {\bibinfo  {journal} {Phys. Rev. Lett.}\ }\textbf {\bibinfo
  {volume} {77}},\ \bibinfo {pages} {3865} (\bibinfo {year}
  {1996})}\BibitemShut {NoStop}%
\bibitem [{\citenamefont {Bl\"ochl}(1994)}]{Blochl1994PBE}%
  \BibitemOpen
  \bibfield  {author} {\bibinfo {author} {\bibfnamefont {P.~E.}\ \bibnamefont
  {Bl\"ochl}},\ }\bibfield  {title} {\bibinfo {title} {{Projector
  augmented-wave method}},\ }\href {https://doi.org/10.1103/PhysRevB.50.17953}
  {\bibfield  {journal} {\bibinfo  {journal} {Phys. Rev. B}\ }\textbf {\bibinfo
  {volume} {50}},\ \bibinfo {pages} {17953} (\bibinfo {year}
  {1994})}\BibitemShut {NoStop}%
\bibitem [{\citenamefont {Marzari}\ and\ \citenamefont
  {Vanderbilt}(1997)}]{marzari1997maximally}%
  \BibitemOpen
  \bibfield  {author} {\bibinfo {author} {\bibfnamefont {N.}~\bibnamefont
  {Marzari}}\ and\ \bibinfo {author} {\bibfnamefont {D.}~\bibnamefont
  {Vanderbilt}},\ }\bibfield  {title} {\bibinfo {title} {{Maximally localized
  generalized Wannier functions for composite energy bands}},\ }\href
  {https://doi.org/10.1103/PhysRevB.56.12847} {\bibfield  {journal} {\bibinfo
  {journal} {Phys. Rev. B}\ }\textbf {\bibinfo {volume} {56}},\ \bibinfo
  {pages} {12847} (\bibinfo {year} {1997})}\BibitemShut {NoStop}%
\bibitem [{\citenamefont {Souza}\ \emph {et~al.}(2001)\citenamefont {Souza},
  \citenamefont {Marzari},\ and\ \citenamefont
  {Vanderbilt}}]{souza2001maximally}%
  \BibitemOpen
  \bibfield  {author} {\bibinfo {author} {\bibfnamefont {I.}~\bibnamefont
  {Souza}}, \bibinfo {author} {\bibfnamefont {N.}~\bibnamefont {Marzari}},\
  and\ \bibinfo {author} {\bibfnamefont {D.}~\bibnamefont {Vanderbilt}},\
  }\bibfield  {title} {\bibinfo {title} {{Maximally localized Wannier functions
  for entangled energy bands}},\ }\href
  {https://doi.org/10.1103/PhysRevB.65.035109} {\bibfield  {journal} {\bibinfo
  {journal} {Phys. Rev. B}\ }\textbf {\bibinfo {volume} {65}},\ \bibinfo
  {pages} {035109} (\bibinfo {year} {2001})}\BibitemShut {NoStop}%
\bibitem [{\citenamefont {Pizzi}\ \emph {et~al.}(2020)\citenamefont {Pizzi},
  \citenamefont {Vitale}, \citenamefont {Arita}, \citenamefont {Bl{\"u}gel},
  \citenamefont {Freimuth}, \citenamefont {G{\'e}ranton}, \citenamefont
  {Gibertini}, \citenamefont {Gresch}, \citenamefont {Johnson}, \citenamefont
  {Koretsune} \emph {et~al.}}]{Pizzi2020wannier90}%
  \BibitemOpen
  \bibfield  {author} {\bibinfo {author} {\bibfnamefont {G.}~\bibnamefont
  {Pizzi}}, \bibinfo {author} {\bibfnamefont {V.}~\bibnamefont {Vitale}},
  \bibinfo {author} {\bibfnamefont {R.}~\bibnamefont {Arita}}, \bibinfo
  {author} {\bibfnamefont {S.}~\bibnamefont {Bl{\"u}gel}}, \bibinfo {author}
  {\bibfnamefont {F.}~\bibnamefont {Freimuth}}, \bibinfo {author}
  {\bibfnamefont {G.}~\bibnamefont {G{\'e}ranton}}, \bibinfo {author}
  {\bibfnamefont {M.}~\bibnamefont {Gibertini}}, \bibinfo {author}
  {\bibfnamefont {D.}~\bibnamefont {Gresch}}, \bibinfo {author} {\bibfnamefont
  {C.}~\bibnamefont {Johnson}}, \bibinfo {author} {\bibfnamefont
  {T.}~\bibnamefont {Koretsune}}, \emph {et~al.},\ }\bibfield  {title}
  {\bibinfo {title} {{Wannier90 as a community code: new features and
  applications}},\ }\href {https://doi.org/10.1088/1361-648x/ab51ff} {\bibfield
   {journal} {\bibinfo  {journal} {J. Phys. Condens. Matter}\ }\textbf
  {\bibinfo {volume} {32}},\ \bibinfo {pages} {165902} (\bibinfo {year}
  {2020})}\BibitemShut {NoStop}%
\bibitem [{\citenamefont {Wu}\ \emph {et~al.}(2018)\citenamefont {Wu},
  \citenamefont {Zhang}, \citenamefont {Song}, \citenamefont {Troyer},\ and\
  \citenamefont {Soluyanov}}]{WU2018WannierTools}%
  \BibitemOpen
  \bibfield  {author} {\bibinfo {author} {\bibfnamefont {Q.}~\bibnamefont
  {Wu}}, \bibinfo {author} {\bibfnamefont {S.}~\bibnamefont {Zhang}}, \bibinfo
  {author} {\bibfnamefont {H.-F.}\ \bibnamefont {Song}}, \bibinfo {author}
  {\bibfnamefont {M.}~\bibnamefont {Troyer}},\ and\ \bibinfo {author}
  {\bibfnamefont {A.~A.}\ \bibnamefont {Soluyanov}},\ }\bibfield  {title}
  {\bibinfo {title} {{WannierTools : An open-source software package for novel
  topological materials}},\ }\href {https://doi.org/10.1016/j.cpc.2017.09.033}
  {\bibfield  {journal} {\bibinfo  {journal} {Comput. Phys. Commun.}\ }\textbf
  {\bibinfo {volume} {224}},\ \bibinfo {pages} {405 } (\bibinfo {year}
  {2018})}\BibitemShut {NoStop}%
\bibitem [{\citenamefont {Zhi}\ \emph {et~al.}(2022)\citenamefont {Zhi},
  \citenamefont {Xu}, \citenamefont {Wu}, \citenamefont {Ning},\ and\
  \citenamefont {Cao}}]{zhi2022wannsymm}%
  \BibitemOpen
  \bibfield  {author} {\bibinfo {author} {\bibfnamefont {G.-X.}\ \bibnamefont
  {Zhi}}, \bibinfo {author} {\bibfnamefont {C.}~\bibnamefont {Xu}}, \bibinfo
  {author} {\bibfnamefont {S.-Q.}\ \bibnamefont {Wu}}, \bibinfo {author}
  {\bibfnamefont {F.}~\bibnamefont {Ning}},\ and\ \bibinfo {author}
  {\bibfnamefont {C.}~\bibnamefont {Cao}},\ }\bibfield  {title} {\bibinfo
  {title} {{WannSymm: A symmetry analysis code for Wannier orbitals}},\ }\href
  {https://doi.org/10.1016/j.cpc.2021.108196} {\bibfield  {journal} {\bibinfo
  {journal} {Comput. Phys. Commun.}\ }\textbf {\bibinfo {volume} {271}},\
  \bibinfo {pages} {108196} (\bibinfo {year} {2022})}\BibitemShut {NoStop}%
\bibitem [{\citenamefont {Merboldt}\ \emph {et~al.}(2025)\citenamefont
  {Merboldt}, \citenamefont {Sch{\"u}ler}, \citenamefont {Schmitt},
  \citenamefont {Bange}, \citenamefont {Bennecke}, \citenamefont {Gadge},
  \citenamefont {Pierz}, \citenamefont {Schumacher}, \citenamefont {Momeni},
  \citenamefont {Steil} \emph {et~al.}}]{merboldt2025observation}%
  \BibitemOpen
  \bibfield  {author} {\bibinfo {author} {\bibfnamefont {M.}~\bibnamefont
  {Merboldt}}, \bibinfo {author} {\bibfnamefont {M.}~\bibnamefont
  {Sch{\"u}ler}}, \bibinfo {author} {\bibfnamefont {D.}~\bibnamefont
  {Schmitt}}, \bibinfo {author} {\bibfnamefont {J.~P.}\ \bibnamefont {Bange}},
  \bibinfo {author} {\bibfnamefont {W.}~\bibnamefont {Bennecke}}, \bibinfo
  {author} {\bibfnamefont {K.}~\bibnamefont {Gadge}}, \bibinfo {author}
  {\bibfnamefont {K.}~\bibnamefont {Pierz}}, \bibinfo {author} {\bibfnamefont
  {H.~W.}\ \bibnamefont {Schumacher}}, \bibinfo {author} {\bibfnamefont
  {D.}~\bibnamefont {Momeni}}, \bibinfo {author} {\bibfnamefont
  {D.}~\bibnamefont {Steil}}, \emph {et~al.},\ }\bibfield  {title} {\bibinfo
  {title} {{Observation of Floquet states in graphene}},\ }\href
  {https://doi.org/0.1038/s41567-025-02889-7} {\bibfield  {journal} {\bibinfo
  {journal} {Nat. Phys.}\ }\textbf {\bibinfo {volume} {21}},\ \bibinfo {pages}
  {1093} (\bibinfo {year} {2025})}\BibitemShut {NoStop}%
\bibitem [{\citenamefont {Sentef}\ \emph {et~al.}(2015)\citenamefont {Sentef},
  \citenamefont {Claassen}, \citenamefont {Kemper}, \citenamefont {Moritz},
  \citenamefont {Oka}, \citenamefont {Freericks},\ and\ \citenamefont
  {Devereaux}}]{sentef2015theory}%
  \BibitemOpen
  \bibfield  {author} {\bibinfo {author} {\bibfnamefont {M.}~\bibnamefont
  {Sentef}}, \bibinfo {author} {\bibfnamefont {M.}~\bibnamefont {Claassen}},
  \bibinfo {author} {\bibfnamefont {A.}~\bibnamefont {Kemper}}, \bibinfo
  {author} {\bibfnamefont {B.}~\bibnamefont {Moritz}}, \bibinfo {author}
  {\bibfnamefont {T.}~\bibnamefont {Oka}}, \bibinfo {author} {\bibfnamefont
  {J.}~\bibnamefont {Freericks}},\ and\ \bibinfo {author} {\bibfnamefont
  {T.}~\bibnamefont {Devereaux}},\ }\bibfield  {title} {\bibinfo {title}
  {Theory of floquet band formation and local pseudospin textures in pump-probe
  photoemission of graphene},\ }\href
  {https://doi.org/https://doi.org/10.1038/ncomms8047} {\bibfield  {journal}
  {\bibinfo  {journal} {Nat. Commun.}\ }\textbf {\bibinfo {volume} {6}},\
  \bibinfo {pages} {7047} (\bibinfo {year} {2015})}\BibitemShut {NoStop}%
\bibitem [{\citenamefont {Freericks}\ \emph {et~al.}(2009)\citenamefont
  {Freericks}, \citenamefont {Krishnamurthy},\ and\ \citenamefont
  {Pruschke}}]{freericks2009theoretical}%
  \BibitemOpen
  \bibfield  {author} {\bibinfo {author} {\bibfnamefont {J.}~\bibnamefont
  {Freericks}}, \bibinfo {author} {\bibfnamefont {H.}~\bibnamefont
  {Krishnamurthy}},\ and\ \bibinfo {author} {\bibfnamefont {T.}~\bibnamefont
  {Pruschke}},\ }\bibfield  {title} {\bibinfo {title} {Theoretical description
  of time-resolved photoemission spectroscopy: Application to pump-probe
  experiments},\ }\href
  {https://doi.org/https://doi.org/10.1103/PhysRevLett.102.136401} {\bibfield
  {journal} {\bibinfo  {journal} {Phys. Rev. Lett.}\ }\textbf {\bibinfo
  {volume} {102}},\ \bibinfo {pages} {136401} (\bibinfo {year}
  {2009})}\BibitemShut {NoStop}%
\bibitem [{\citenamefont {Farrell}\ \emph {et~al.}(2016)\citenamefont
  {Farrell}, \citenamefont {Arsenault},\ and\ \citenamefont
  {Pereg-Barnea}}]{farrell2016dirac}%
  \BibitemOpen
  \bibfield  {author} {\bibinfo {author} {\bibfnamefont {A.}~\bibnamefont
  {Farrell}}, \bibinfo {author} {\bibfnamefont {A.}~\bibnamefont {Arsenault}},\
  and\ \bibinfo {author} {\bibfnamefont {T.}~\bibnamefont {Pereg-Barnea}},\
  }\bibfield  {title} {\bibinfo {title} {Dirac cones, floquet side bands, and
  theory of time-resolved angle-resolved photoemission},\ }\href
  {https://doi.org/https://doi.org/10.1103/PhysRevB.94.155304} {\bibfield
  {journal} {\bibinfo  {journal} {Phys. Rev. B}\ }\textbf {\bibinfo {volume}
  {94}},\ \bibinfo {pages} {155304} (\bibinfo {year} {2016})}\BibitemShut
  {NoStop}%
\end{thebibliography}%

\end{document}